\documentclass[longauth]{aa}

\usepackage{graphicx}
\usepackage{multirow}
\usepackage{xcolor}
\usepackage{txfonts}
\begin{document} 

   \title{Single-lens mass measurement in the high-magnification microlensing event Gaia19bld located in the Galactic disc}

   \author{K. A. Rybicki,\thanks{krybicki@astrouw.edu.pl}
          \inst{\ref{oauw}}
          {\L}. Wyrzykowski,
          \inst{\ref{oauw}}
          E. Bachelet,
          \inst{\ref{lco}}
          A. Cassan,
          \inst{\ref{iap}}
          P. Zieli{\'n}ski,
          \inst{\ref{oauw}}
          A. Gould,
          \inst{\ref{max}, \ref{osu}}
          S. Calchi Novati,
          \inst{\ref{ipac}}
          J.C. Yee,
          \inst{\ref{cfa}}
          Y.-H. Ryu,
          \inst{\ref{Korean}}
          M. Gromadzki,
          \inst{\ref{oauw}}
          P. Miko{\l}ajczyk,
          \inst{\ref{uwr}}
          N. Ihanec,
          \inst{\ref{oauw}}
          K. Kruszy{\'n}ska,
          \inst{\ref{oauw}}
          F.-J. Hambsch,
          \inst{\ref{road}, \ref{Josch2}}
           S. Zo{\l}a,
          \inst{\ref{uj}}
          S. J. Fossey, 
          \inst{\ref{ucl}}
          S. Awiphan,
          \inst{\ref{Thai1}}
          N. Nakharutai,
          \inst{\ref{Thai2}}
          F. Lewis,
          \inst{\ref{faulkes1}, \ref{faulkes2}}
          F. Olivares E.,
          \inst{\ref{UdeChile}}
          S. Hodgkin,
          \inst{\ref{cam}}
          A. Delgado,
          \inst{\ref{cam}}
          E. Breedt,
          \inst{\ref{cam}}
          D. L. Harrison,
          \inst{\ref{cam}, \ref{kavli}}            
          M. van Leeuwen,
          \inst{\ref{cam}}
          G. Rixon,
          \inst{\ref{cam}}
          T. Wevers,
          \inst{\ref{cam}}
          A. Yoldas,
          \inst{\ref{cam}}         
          A. Udalski,
          \inst{\ref{oauw}}
          M. K. Szyma{\'n}ski,
          \inst{\ref{oauw}}
          I. Soszy{\'n}ski,
          \inst{\ref{oauw}}
          P. Pietrukowicz,
          \inst{\ref{oauw}}
          S. Koz{\l}owski,
          \inst{\ref{oauw}}
          J. Skowron,
          \inst{\ref{oauw}}
          R. Poleski,
          \inst{\ref{oauw}}
          K. Ulaczyk,
          \inst{\ref{warwick},\ref{oauw}}
          P. Mr{\'o}z,
          \inst{\ref{oauw},\ref{caltech}}
          P. Iwanek,
          \inst{\ref{oauw}}
          M. Wrona,
          \inst{\ref{oauw}}
          R.A. Street,
          \inst{\ref{lco}}
          Y. Tsapras,
          \inst{\ref{heidel}}
          M. Hundertmark,
          \inst{\ref{heidel}}
          M. Dominik,
          \inst{\ref{stAndrews}}          
          C. Beichman,
          \inst{\ref{ipac}}
          G. Bryden,
          \inst{\ref{ipac}}          
          S. Carey,
          \inst{\ref{ipac}}
          B.S. Gaudi,
          \inst{\ref{osu}}
          C. Henderson,
          \inst{\ref{ipac}}          
          Y. Shvartzvald,
          \inst{\ref{weizman}}
          W. Zang,
          \inst{\ref{Tsinghua}}            
          W. Zhu,
          \inst{\ref{Toronto}}
          G.W. Christie,
          \inst{\ref{AA}}
          J. Green,
          \inst{\ref{AB}}
          S. Hennerley,
          \inst{\ref{AB}}
          J. McCormick,
          \inst{\ref{AC}}
          L.A.G. Monard,
          \inst{\ref{AD}}
          T. Natusch,
          \inst{\ref{AA},\ref{AE}}
          R.W. Pogge,
          \inst{\ref{osu}}
          I. Gezer,
          \inst{\ref{oauw}}
          A. Gurgul,
          \inst{\ref{oauw}}
          Z. Kaczmarek,
          \inst{\ref{oauw}}
          M. Konacki,
          \inst{\ref{camk}}
          M. C. Lam,
          \inst{\ref{oauw},\ref{faulkes2}}
          M. Maskoliunas,
          \inst{\ref{Vilnus}}
          E. Pakstiene,
          \inst{\ref{Vilnus}}
          M. Ratajczak,
          \inst{\ref{oauw}}
          A. Stankeviciute,
          \inst{\ref{oauw}}
          J. Zdanavicius,
          \inst{\ref{Vilnus}}
          O. Zi{\'o}{\l}kowska
          \inst{\ref{oauw}}
          }
   \institute{Astronomical Observatory, University of Warsaw, Al. Ujazdowskie 4, 00-478, Warszawa, Poland
   \label{oauw}
        \and
            Las Cumbres Observatory, 6740 Cortona Drive, suite 102, Goleta, CA 93117, USA
            \label{lco}
        \and
         Institut d'Astrophysique de Paris, Sorbonne Universit\'e, CNRS, UMR 7095, 98 bis bd Arago, F-75014 Paris, France
        \label{iap}
        \and
            Max Planck Institute for Astronomy, Konigstuhl 17, 69117 Heidelberg, Germany
            \label{max}
        \and
            Department of Astronomy, Ohio State University, 140 West 18th Avenue, Columbus, OH 43210, USA
            \label{osu}
        \and
            IPAC, Mail Code 100-22, Caltech, 1200 E. California Blvd., Pasadena, CA 91125, USA
            \label{ipac}
        \and
            Center for Astrophysics $|$ Harvard \& Smithsonian, 60 Garden St, MS-15 Cambridge, MA 02138, USA
            \label{cfa}
        \and
            Korea Astronomy and Space Science Institute, Daejon 34055, Republic of Korea     
            \label{Korean}
        \and
            Astronomical Institute, University of Wroc{\l}aw, ul. Kopernika 11, 51-622, Wroc{\l}aw, Poland
            \label{uwr}            
        \and        
            ROAD Observatory, San Pedro de Atacama, Chile
            \label{road}
        \and
            Vereniging Voor Sterrenkunde (VVS), Oostmeers 122 C, 8000 Brugge, Belgium
            \label{Josch2}
        \and
            Astronomical Observatory of the Jagiellonian University, ul. Orla 171, 30-244 Kraków, Poland
            \label{uj}
        \and
            Dept. of Physics and Astronomy, University College London, Gower St., London WC1E 6BT, UK
            \label{ucl}
        \and
            National Astronomical Research Institute of Thailand, 260 Moo 4, Donkaew, Mae Rim, Chiang Mai 50180, Thailand
            \label{Thai1}
        \and
            Data Science Research Center, Department of Statistics, Faculty of Science, Chiang Mai University, Chiang Mai 50200, Thailand
            \label{Thai2}
        \and
            Faulkes Telescope Project, School of Physics and Astronomy, Cardiff University, The Parade, Cardiff, CF24 3AA, Wales, UK
            \label{faulkes1}
        \and
            Astrophysics Research Institute, Liverpool John Moores University, 146 Brownlow Hill, Liverpool L3 5RF, UK
            \label{faulkes2}
        \and
            Instituto de Astronom\'{\i}a y Ciencias Planetarias, Universidad de Atacama, Copayapu 485, Copiap\'o, Chile
            \label{UdeChile}
        \and    
            Institute of Astronomy, University of Cambridge, Madingley Road, Cambridge CB3 0HA, UK
            \label{cam}
        \and
            Kavli Institute for Cosmology Cambridge, Institute of Astronomy, Madingley Road, Cambridge, CB3 0HA, UK
            \label{kavli}
        \and
            Department of Physics, University of Warwick, Coventry CV4 7 AL, UK
            \label{warwick}
        \and
            Division of Physics, Mathematics, and Astronomy, California Institute of Technology, Pasadena, CA 91125, USA
            \label{caltech}
        \and
            Zentrum f{\"u}r Astronomie der Universit{\"a}t Heidelberg, Astronomisches Rechen-Institut, M{\"o}nchhofstr. 12-14, 69120 Heidelberg, Germany
            \label{heidel}
        \and
            University of St Andrews, College Gate, St Andrews KY16 9AJ, UK
            \label{stAndrews}
        \and
            Department of Particle Physics and Astrophysics, Weizmann Institute of Science, Rehovot 76100, Israel
            \label{weizman}
        \and
            Physics Department and Tsinghua Centre for Astrophysics, Tsinghua University, Beijing 100084, People's Republic of China
            \label{Tsinghua}
        \and
            Canadian Institute for Theoretical Astrophysics, University of Toronto, 60 St. George Street, Toronto, ON M5S 3H8, Canada
            \label{Toronto}
        \and
            Auckland Observatory, Auckland, New Zealand
            \label{AA}
        \and
            Kumeu Observatory, Kumeu, New Zealand
            \label{AB}
        \and
            Farm Cove Observatory, Centre for Backyard Astrophysics, Pakuranga, Auckland, New Zealand
            \label{AC}
        \and
            Klein Karoo Observatory, Centre for Backyard Astrophysics, Calitzdorp, South Africa
            \label{AD}
        \and
            Institute for Radio Astronomy and Space Research (IRASR), AUT University, Auckland, New Zealand
            \label{AE}
        \and
            Nicolaus Copernicus Astronomical Center, Polish Academy of Sciences, Bartycka 18, 00-716 Warsaw, Poland
            \label{camk}
        \and
            Institute of Theoretical Physics and Astronomy, Vilnius University, Saulėtekio al. 3, Vilnius, LT-10257, Lithuania
            \label{Vilnus}
            }
   \date{}


  \abstract
{Microlensing provides a unique opportunity to detect non-luminous objects. In the rare cases that the Einstein radius $\theta_{\rm E}$ and microlensing parallax $\pi_{\rm E}$ can be measured, it is possible to determine the mass of the lens. With technological advances in both ground- and space-based observatories, astrometric and interferometric measurements are becoming viable, which can lead to the more routine determination of $\theta_{\rm E}$ and, if the microlensing parallax is also measured, the mass of the lens.}
   {We present the photometric analysis of Gaia19bld, a high-magnification ($A\approx60$) microlensing event located in the southern Galactic plane, which exhibited finite source and microlensing parallax effects. Due to a prompt detection by the Gaia satellite and the very high brightness of $I = 9.05~$mag at the peak, it was possible to collect a complete and unique set of multi-channel follow-up observations, which allowed us to determine all parameters vital for the characterisation of the lens and the source in the microlensing event.}
   {Gaia19bld was discovered by the Gaia satellite and was subsequently intensively followed up with a network of ground-based observatories and the Spitzer Space Telescope. We collected multiple high-resolution spectra with Very Large Telescope (VLT)/X-Shooter to characterise the source star. The event was also observed with VLT Interferometer (VLTI)/PIONIER during the peak. Here we focus on the photometric observations and model the light curve composed of data from Gaia, Spitzer, and multiple optical, ground-based observatories. We find the best-fitting solution with parallax and finite source effects.
   We derived the limit on the luminosity of the lens based on the blended light model and spectroscopic distance.}
   {
   We compute the mass of the lens to be $1.13 \pm 0.03~M_{\odot}$ and derive its distance to be $5.52^{+0.35}_{-0.64}~\mathrm{kpc}$.
   The lens is likely a main sequence star, however its true nature has yet to be verified by  future high-resolution observations.

   Our results are consistent with interferometric measurements of the angular Einstein radius, emphasising that interferometry can be a new channel for determining the masses of objects that would otherwise remain undetectable, including stellar-mass black holes.}
   {}

   \titlerunning{Mass measurement of the lens in Gaia19bld}
   \authorrunning{Rybicki K. A. et al.}
   
   \maketitle
   
    \section{Introduction}
     
    Determining the masses of stars in our Galaxy is not an easy task for astronomers. Even though evolutionary models of stars can give us rough estimates of their masses, actual measurements are necessary to confirm theoretical predictions, which is crucial for studies of the structure of the Galaxy, the late stages of stellar evolution, and the distribution and mass function of stellar remnants in the Milky Way. 
    
    Measuring the mass of a star or a stellar remnant usually requires measuring its gravitational interactions with the local environment. In the case of multiple systems, it is the dynamical interaction with the companion star, but techniques differ depending on the type of the system. In the simplest scenario, one can resolve images of the objects and simply measure their projected orbits over the course of a few years of observations, assuming the orbital period is short enough (e.g. \citealt{LeBouquin2017, Gillessen2012}).  This is a rather rare scenario, and much more commonly dynamical mass measurements are made using the radial velocity technique (e.g. \citealt{Graczyk2018}), which still requires us to detect the light from visible component(s).
    Another still developing, although already well-established method of determining the masses of stars is asteroseismology (e.g. \citealt{Kjeldsen1995}). Its main limitation is the requirement of high-precision and often also high-cadence photometry. The CoRoT \citep{Kallinger2010} and Kepler \citep{Gilliland2010} missions were revolutionary for the field and allowed the measurement of stellar parameters (including masses) of thousands of stars. It is worth noting here that microlensing of a star with measured asteroseismc oscillations can be highly beneficial because it unravels precious information about the source \citep{Li2019}. This synergy between microlensing and asteroseismology has a chance to be further exploited with NASA Roman Space Telescope (formerly known as the Wide Field Infrared Survey Telescope, WFIRST, launch planned for $\sim$ 2025), which will provide a high-cadence photometric time series for a region located in the Galactic plane \citep{Penny2019}.
    For more exotic systems there are other ways of measuring the mass, for example  the pulsar timing method for young neutron stars (NS) and pulsars (e.g. \citealt{Kiziltan2013}) and gravitational wave detection for coalescent binary black hole (BH) systems (e.g. \citealt{Abbott2016}), NS+NS \citep{Abbott2017Multi, Smartt2017Nature}, or BH+NS (e.g. \citealt{Ackley2020}).
    While each of the mentioned techniques can be useful in a specific scenario, none of them will suffice when dealing with isolated faint (or dark) stellar remnants, which are of particular interest to evolutionary models.
    As isolated BHs are especially difficult to detect, many questions regarding them remain unanswered: How many of them are there in our Galaxy? What is their mass function? Is there a mass gap between NSs and BHs \citep{Ozel2010, Wyrzykowski2016, WyrzykowskiMandel2020, Olejak2020}? Thanks to the gravitational microlensing technique, these questions may find answers in the near future.
    
    Gravitational microlensing phenomena \citep{1936Einstein, Liebes1964, Refsdal964, Paczynski1986} have been observed and used for various applications for almost 30 years.  The most prominent effect of microlensing is the increase in brightness of a background star (`source') due to the presence of a mass that is passing between the source and the observer, acting as a `lens'. The advantage of microlensing in the context of mass measurement methods is that the only light needed for an event to occur is the light from the background source star;  indeed, the lens light can be considered as `contaminating' and can slightly complicate the calculations and the modelling process. This makes observations of microlensing phenomena the natural and only viable method to detect and measure masses of isolated dark objects like black holes or neutron stars.
    
    The downside is that, in a typical event, the brightness change alone does not provide enough information about the physical properties of the lens and the source. This is because the distance to the source $D_{\rm S}$, distance to the lens $D_{\rm L}$, and the mass of the lens $M_{\rm L}$, as well as the lens--source relative proper motion $\mu_{\rm rel}$ are degenerated into a single parameter, the Einstein time $t_{\rm E}$, which is the time needed for the relative position of the lens and the source to change by the angular Einstein radius $\theta_{\rm E}$:
    \begin{align}
        t_{\rm E} = \frac{\theta_{\rm E}}{\mu_{\rm rel}}, && \theta_{\rm E}=\sqrt{\kappa M_{\rm L} \pi_{\rm rel}}, && \pi_{\rm rel} = \frac{1~\mathrm{au}}{D_{\rm L}} - \frac{1~\mathrm{au}}{D_{\rm S}}.
    \label{eq_theta}    
    \end{align}
    Here $\mu_{\rm rel}$ is the proper motion of the lens relative to the source and $\kappa = 8.144~\frac{\rm mas}{M_{\odot}}$. The Einstein radius $\theta_{\rm E}$ is the theoretical radius of the ring-shaped image that would form in an ideal case of the perfect alignment of the point source, the point lens, and the observer.
    
    In some events it is possible to  measure the so-called microlensing parallax vector parameter $\vec{\pi_{\rm E}}$, which is related to the actual geometric parallax through the Einstein radius \citep{Gould1992, Gould2004}:
    \begin{equation}
        \vec{\pi_{\rm E}} = \frac{\pi_{\rm rel}}{\theta_{\rm E}} \frac{\vec{\mu_{\rm rel}}}{\mu_{\rm rel}}
    \label{eq_piE}        
    .\end{equation}
    Formulae (1) and (2) immediately show that
    \begin{align}
        M_{\rm L} = \frac{\theta_{\rm E}}{\kappa \pi_{\rm E}}, &&
        D_{\rm L} = \left( D_{\rm S}^{-1} + \frac{\pi_{\rm E}\theta_{\rm E}}{\rm au} \right)^{-1} \label{eq_mass} .   
    \end{align}
    The mass formula is particularly important here;  it implies that for events with measured parallax effect the only quantity required to derive the lens mass is $\theta_{\rm E}$.

    There are two ways to determine the Einstein radius from the photometry of microlensing events, but both of them are challenging and require a special geometry. The first and most straightforward approach is to measure relative proper motion by directly detecting the light from the lens, once the lens and the source are well separated (e.g. \citealt{Alcock2001}). This yields $\theta_{\rm E}$, under the assumption that the $t_{\rm E}$ is measured.
    This method is becoming more viable now as microlensing observations span almost 30 years, and so the number of events with separated sources and lenses increases. The Einstein radius has been measured a few times using this procedure (e.g. \citealt{Bennett2015}, \citealt{Bhattacharya2018}, \citealt{Bennett2020}, \citealt{Vandorou2020}), but these measurements are still challenging to perform. However, this method is likely to become more routine in the era of 30 m adaptive optics (AO) roughly one decade from now.
    In addition, it  requires the lens to be luminous, which makes it unusable for detecting dark stellar remnants. The second opportunity to measure $\theta_{\rm E}$ arises thanks to finite source effects (\citealt{Gould1994}, \citealt{WittMao1994}, \citealt{Nemiroff1994}) in high-magnification  single-lens events or in binary-lens events, when the source crosses the caustic (see   section 2 for more details).
    
    Another channel to measure $\theta_{\rm E}$ is from astrometry. With the increasing precision of astrometric measurements, the positional methods of observing microlensing events are becoming viable. The effect of the displacement of the light centroid, referred to as astrometric microlensing (\citealt{Hog1995}, \citealt{Miyamoto1995}, \citealt{Walker1995}, \citealt{Dominik2000}), has already been measured twice \citep{Sahu2017, Zurlo2018}, but given that we are at the beginning of the era of advanced space satellite missions, such measurements may become more and more common. The Gaia satellite will likely provide several microlensing events with detectable astrometric signals \citep{Rybicki2018, Kluter2019}, and with the Roman Space Telescope, the lens masses could be measured regularly, rather than for the very special cases, allowing us to probe the population of black holes and other stellar remnants, as well as regular stars \citep{GouldYee2014}.
    
    Last but not least, it is possible to directly measure the $\theta_{\rm E}$ parameter by resolving the images with high-precision interferometers \citep{Delplancke2001, Cassan2016}. Because it requires very bright targets, successful observations have been conducted only twice so far. For the Kojima event (\citealt{Nucita2018, Fukui2019, Zang2020}) the interferometric measurements led to the separation of images using the Gravity instrument on ESO VLTI \citep{Dong2019}. The second  was the Gaia19bld event, for which the photometric analysis is presented in this paper, while the interferometric measurements are described in companion paper (\citealt{Cassan2021}, hereafter C21).
    
    Here we report on the detection and extensive photometric follow-up of a high-magnification microlensing event Gaia19bld found in the Galactic disc, exhibiting finite-source effects in the light curve, as well as a strong detectable microlensing parallax signal. 
    Most of the events known to date were found in dedicated surveys, especially Optical Gravitational Lensing Experiment (OGLE, \citealt{Udalski1992, Udalski2015}), Microlensing Observations in Astrophysics (MOA, \citealt{MOA2001}), and Korea Microlensing Telescope Network (KMTNet, \citealt{Kim2016}). These projects focus on the monitoring of the Galactic bulge because the optical depth for detecting an event is the highest towards the centre of our Galaxy. For most of the sources in such events it is possible to estimate $D_{\rm S}$ from the colour-magnitude diagram. This is usually not possible for  disc events, due to the lack of red-clump stars, and then spectroscopic methods are required (e.g. \citealt{WyrzykowskiGaia16aye}).
    Gaia19bld is the unique case of a microlensing event for which almost all of the interesting physical parameters of the source and the lens have been derived, thanks to an observing strategy that draws on multiple types of follow-up observations.
    
    The event was detected by the Gaia satellite \citep{Gaia19bldTNS, Gaia19bldATel} and observed intensively with a network of ground-based telescopes. Thanks to dense photometric monitoring, it was possible to predict the time and magnification at the peak, when the event was bright enough to be observed with VLTI/PIONIER. This allowed, for the first time ever, to detect the motion of the microlensing images (see C21). The event was observed with high-resolution spectrographs, the  VLT/X-Shooter and the Las Cumbres NRES, to further characterise the source and the lens (\citealt{Bachelet2021}, hereafter B21). Finally, the follow-up observations from the Spitzer Space Telescope allowed  the precise measurement of the microlensing parallax signal. The case of Gaia19bld shows the great potential of combining photometric, astrometric, spectroscopic, and interferometric follow-up observations of microlensing events to characterise lens and source stars. In particular, the accurate derivation of lens masses, even for dark objects like neutron stars or black holes.
   
    The aim of this paper,  one in a series of three publications on the Gaia19bld event (see C21 and B21), is to show the photometric part of the analysis and is organised as follows. In Sect. 2 we briefly summarise the basics of microlensing, including the necessary formulae and explanations of the second-order effects. In Sect. 3 we present the observing strategy and data reduction process. Section 4 contains the description of the microlensing model and the challenges related to the modelling process. We derive physical parameters of the source and lens in Sect.  5, discuss our results in Sect.  6, and provide the summary in Sect. 7.
    
    \section{Microlensing essentials}
    Microlensing occurs when the observer, the lensing star, and a background source are almost perfectly co-linear. Because stars in the Galaxy are in constant motion, the   configuration is not static:  the magnification $A(t)$ changes with the projected separation $u(t)$ of the lens and the source. For the standard microlensing event (a point source and a  point lens) it can be described as \citep{Paczynski1986}:
    \begin{align}
        A(t)=\frac{u(t)^2 + 2}{u(t)\sqrt{u(t)^2+4}}, && u(t)=\sqrt{u_{\rm 0}^2+\tau(t)^2}, && \tau(t)=\frac{t-t_{\rm 0}}{t_{\rm E}}.
    \label{eq_A}    
    \end{align}
    Here $u_0$ is the smallest projected separation of the source and the lens (in units of $\theta_{\rm E}$), when the magnification is the highest, while $t_0$ is the time of the maximum, so that $u(t_0)=u_0$.
    The total flux during the microlensing event can be written as
    \begin{align}
        F_{\rm tot}(t) = F_{\rm S} A(t) + F_{\rm bl},
    \end{align}
    where $F_{\rm S}$ is the flux from the source and $F_{\rm bl}$ is the flux from the blend, which is a combined flux from the lens and background stars aligned with the line of sight and blended with the source (unresolved). Because three parameters ($t_0$, $t_{\rm E}$, $u_0$) are needed to compute the magnification $A(t)$ and two more ($F_{\rm S}$ and $F_{\rm bl}$) for the description of total flux changes, five parameters define a microlensing light curve in the simplest scenario, where both lens and source are considered to be single point-sized objects.
    
    The important implication of Formulae \ref{eq_theta}-\ref{eq_mass} is that the Einstein time $t_{\rm E}$ is the only parameter connected to the physical properties of the lens that can be derived from the light curve of the standard event. In Gaia19bld there are two prominent second-order effects detectable in the light curve, the finite size of the source and the microlensing parallax signal, which allowed for almost complete characterisation of the lens from the photometric measurements alone.
    
    \subsection{Annual parallax}
    
    Earth's orbital motion causes changes in the observer's position relative to the source direction, which in consequence leads to changes in the lens-source projected separation and observed magnification. Thus, the projected separation definition requires modification
    \begin{align}
      u(t)=\sqrt{\beta(t)^2 + \tau(t)^2}, && \tau = \frac{t-t_0}{t_{\rm E}} + \delta \tau, && \beta = u_{\rm 0} + \delta \beta,
    \end{align}
    where $(\delta \tau, \delta \beta)$ is the displacement vector of the projected relative position of the source and the lens due to the parallax.
    While this effect is always present, it is only detectable for a  handful of events, preferentially those that last long enough for  Earth to substantially change its position on the orbit. We adopt the geocentric frame as described by \cite{Gould2004} and introduce the microlensing parallax vector $\vec{\pi_{\rm E}}$ \citep{Gould2000b},
    which is defined so that its direction is the same as that of the proper motion of the lens relative to the source, and its magnitude is the relative parallax of the source and the lens scaled by the Einstein radius (Formula \ref{eq_piE}). The microlensing parallax vector $\vec{\pi_{\rm E}}$ serves to define the time-dependant microlensing parallax shift as
    \begin{equation}
        (\delta \tau, \delta \beta) = (\vec{\pi_{\rm E}} \cdot \vec{\Delta s}, \vec{\pi_{\rm E}} \times \vec{\Delta s} ),
    \end{equation}
    where $\delta \tau$ is the component towards the relative motion of the source and the lens, $\delta \beta $ is perpendicular to it, and the vector \vec{\Delta s} is a positional offset of the Sun projected onto the sky.
   \subsection{Space parallax}
    Another well-developed method to measure the $\vec{\pi_{\rm E}}$ vector is the so-called space parallax effect. Instead of continuous observations from Earth as it changes position on its orbit, measuring space parallax requires registering the light curve from two different observatories separated by a substantial fraction of the Einstein ring projected on the observer plane, meaning that the second observatory has to be a space satellite, located roughly $\sim 1~ \rm au$ from Earth. As the lens-source projected separation registered by ground-based and space-based observatories will be different, the microlensing light curves will also differ. To first order, by comparing microlensing parameters measured from space ($t_{0, \rm sat}$, $t_{\rm E, \rm sat}$, $u_{0, \rm sat}$) and from the ground, we can derive the microlensing parallax vector as (\citealt{Refsdal1966}, \citealt{Gould1994_space})
    \begin{align}
      \vec{\pi_{\rm E}}=\frac{\rm au}{D_{\perp}}(\Delta \tau, \Delta u_0), && \Delta \tau = \frac{t_{0, \rm sat} - t_0}{t_{\rm E}}, && \Delta u_0 = u_{0, \rm sat} - u_0,
    \end{align}
    where $D_{\perp}$ is the distance between Earth and the satellite projected on the sky. In practice, when solving for
    the space parallax, we take into account the full position and motion information for the spacecraft as a function of time. 
    \subsection{Finite source effect}
    
    If the projected separation of the source and the lens is comparable to the angular radius of the source $\theta_*$, we can no longer use the formula for magnification as presented in Formula \ref{eq_A}, but we have to integrate over the whole area of the source disc $S_{\rm src}$. Then the magnification can be written as
    \begin{align}
    A_{\rm FS}(t,\rho) = \frac{1}{\pi \rho^2}\int_{S_{\rm src}} A(t)\,dS_{\rm src}, && \rho=\frac{\theta_*}{\theta_E}.
    \label{eq_rho}
    \end{align}
    The finite source effect is prominent especially for events, for which $u_0 < \rho$. For such events we are essentially dealing with caustic crossing (the caustic is the point in this case) because the lens is crossing in front of the  very disc of the source star. If the light curve is sampled well enough during the peak of the event, the $\rho$ parameter can be obtained. Measuring the angular radius of the source $\theta_*$ is possible using information from spectroscopy or empirical colour-angular diameter relations (e.g. \citealt{Adams2018}). Having these two parameters allows  $\theta_{\rm E}$ to be derived (Formula \ref{eq_rho}), which in turn can serve to determine important physical parameters of the lens, namely its mass, distance, and transverse velocity.  

    \section{Discovery and follow-up data}
    
    The Gaia19bld transient (AT 2019dqb at the IAU Transient Name Server), located at RA$_{\mathrm{J}2000}$=12:37:32.56, Dec$_{\mathrm{J}2000}$=-66:06:40.90, which corresponds to the Galactic coordinates $l=301^{\circ}.52358$, $b=-3^{\circ}.27762$, was discovered and flagged by Gaia Science Alerts\footnote{http://gsaweb.ast.cam.ac.uk/alerts/alert/Gaia19bld} (\citealt{Wyrzykowski2012, Hodgkin2013, Hodgkin2021}) on 18 April 2019 as a `long-term rise of $\sim$0.4 magnitudes in a bright Gaia source'. It was tentatively classified as a candidate microlensing event because there was no prior variability in the Gaia $G$-band light curve, and BP-RP low-resolution spectra did not show any specific emission lines or spectral evolution that are often connected with other types of transient objects. In addition, the object was alerted relatively early, so the rising part of the light curve could be covered. This is not always possible for alerts provided by Gaia because of the low cadence of observations, especially in regions near the ecliptic, which includes the Galactic bulge. Thanks to the high brightness of the event, it could be easily followed up  even with smaller telescopes, which allowed for dense coverage from the ground. The data contributed by all observatories participating in this follow-up campaign are summarised in Table \ref{tab_all_obs}.
    
    \subsection{Gaia data}
    
    Gaia19bld was alerted at the beginning of the rise, about 0.4 mag above the baseline brightness, which was then revealed to be $G\approx14.8~\mathrm{mag}$.  Due to the Gaia scanning law, there are usually two measurements separated by 6 hours, and the same object is observed again after an average of about one month. For this particular field, the Gaia cadence was adequate to sample the baseline, but only 12 Gaia epochs (each consisting of two individual measurements) were collected during the event. Thus, the exact shape of the light curve, parallax effect and finite source effect could not be measured without an intensive follow-up campaign. Although the Gaia Science Alerts web page does not provide uncertainties of the individual measurements, the nominal photometric error for Gaia19bld should vary from $0.006~\mathrm{mag}$ for $G=14.8~\mathrm{mag}$ at the baseline to $0.003~\mathrm{mag}$ for $G=10.50~\mathrm{mag}$ at the peak  \citep{GaiaDR2photo}.
    
    \subsection{Photometric follow-up}
    The Gaia Science Alerts Follow-up Network is an informal group of astronomers, both professionals and amateurs, coordinated via an OPTICON EC grant\footnote{www.astro-opticon.org}. Members of this network have access to telescopes of sizes ranging from 25cm to 2m (mostly 1m class telescopes) and are willing to observe alerts provided by Gaia and other brokers. This collaboration has already produced multiple interesting discoveries (e.g. \citealt{Campbell2015, WyrzykowskiGaia16aye, Szegedi-Elek2020}).
    \begin{table}[h]
      \centering
      \caption[]{Photometric data collected for Gaia19bld.}
      \label{tab_all_obs}

         \begin{tabular}{ccc}
            \hline
            \hline
            \noalign{\smallskip}
            Observatory & Filters & Data points \\
            \noalign{\smallskip}
            \hline
            \noalign{\smallskip}
            Gaia & $G$ & 82 \\
            Spitzer & 3.6$~\mu$m & 35 \\
            OGLE & $I$ & 217 \\
            \hline
            LCOGT SSO: \\
            \hline
            coj1m003 & $I,V$ & 206 \\
            coj1m011 & $I,V$ & 321 \\
            \hline
            LCOGT CTIO: \\
            \hline
            lsc1m004 & $I,V$ & 297 \\
            lsc1m005 & $I,V$ & 254 \\
            lsc1m009 & $I,V$ & 20 \\
            \hline
            LCOGT SAAO: \\
            \hline
            cpt1m010 & $I,V$ & 170 \\
            cpt1m012 & $I,V$ & 262 \\
            cpt1m013 & $I,V$ & 302 \\
            ROAD & $V$ & 3211 \\
            \hline
            $\mu$FUN: \\
            \hline
            Kumeu & $R$ & 243\\
            KKO & $R$ & 5517\\
            AO & $R$ & 456\\
            FCO & - & 322\\
            CT13 & $I$ & 545\\
            \hline
            SKYNET & $i', r', V$ & 5815 \\
            Telescope Live AUS-1 & $r', V$ & 131 \\
            Telescope Live CHI-1 & $r', g'$ & 39 \\
            REM & $i', r', g'$ & 30 \\
            \hline
            \hline
         \end{tabular}

   \end{table}
    For the case of Gaia19bld, the most important data were collected using the Las Cumbres Observatory global network (LCOGT) of robotic telescopes. Because the event is located in the southern hemisphere, we used the network sites at the Cerro Tololo Inter-American Observatory (CTIO, $30^{\circ}10'11''\mathrm{S},~70^{\circ}48'23''\mathrm{W}$), the Siding Spring Observatory (SSO, $31^{\circ}16'24''\mathrm{S},~149^{\circ}3'52''\mathrm{E}$), and the South African Astronomical Observatory (SAAO, $32^{\circ}22'34''\mathrm{S},~20^{\circ}48'38''\mathrm{E}$) located in Chile, Australia, and South Africa, respectively. Thanks to the quick response of the robotic telescopes, observations taken using these telescopes were the first follow-up taken from the ground (April 23, four days after the alert) and they cover the whole duration of the event, including the most important part at the peak. They were reduced in almost real time to ensure the most precise prediction of the time of the maximum for the interferometric follow-up (see more in C21). All the measurements were taken using the Sinistro imagers in $V$ and $I$ Johnson-Cousins filters to match available catalogues. For all Las Cumbres data, the reduction was done using CCDPhot package (Mikolajczyk et al. in prep),  a flexible tool designed to perform precise photometry and astrometry on multi-site imaging data. It uses SExtractor and Scamp \citep{sextractor1996, scamp2006} for the initial photometry and astrometry, but the final reduction process is based mostly on the Daophot package \citep{Stetson1987} and the Pyraf package.
    
    The second follow-up data set was collected using a 40cm robotic telescope in the ROAD observatory by an amateur astronomer Dr. Franz-Josef (Josch) Hambsch, an active participant of the Gaia Science Alerts Follow-up Network. The observatory is located near San Pedro de Atacama ($22^{\circ}57'10''\mathrm{S},~68^{\circ}10'49''\mathrm{W}$), and it consists of multiple smaller telescopes, mostly used by amateurs, who often collaborate with professional astronomers. Observations from this telescope provided a complete and homogeneous data set. Even though the single-epoch errors are rather large for all but the brightest parts of the light curve, the number of images taken is enough to provide sufficient photometric accuracy. All images of Gaia19bld from this site were taken in V band, and the standard photometry was extracted using LesvePhotometry software.
    \begin{figure}[t]
        \centering
        \includegraphics[width=0.9\columnwidth]{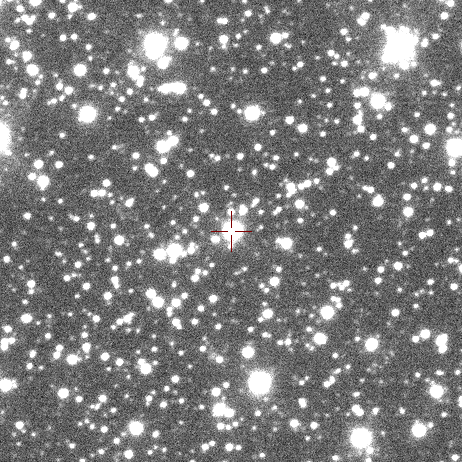}
        \caption{Section of the OGLE-IV reference image (2x2 arcmin). The Gaia19bld source is shown as a red cross. It shows the neighbourhood of the star prior to the event;  the image is a composition of ten $I$-band frames taken between 5 Feb 2014 and 30 Jan 2015. North is up; east is to the left.}
        \label{fig_ogle_zoom}
    \end{figure}
    
    High-cadence photometric follow-up during the peak of the event was provided by five telescopes that are members of the $\mu$FUN network. Three of the instruments are located in New Zealand,   the 0.41 m telescope at the Kumeu Observatory ($36^{\circ} 48'24''\mathrm{S},~174^{\circ}31'29"\mathrm{E}$), the 0.4 m telescope at the  Auckland Observatory (AO, $174^{\circ} 46'37''\mathrm{E},~36^{\circ} 54'22''\mathrm{S}$), and the  0.36 m telescope at the  Farm Cove Observatory (FC, $174^{\circ} 53'37''\mathrm{E},~36^{\circ}53'37''\mathrm{S}$); one  in South Africa,  the 
0.36 m telescope at the  Klein Karoo Observatory (KK, $21^{\circ}40'00''\mathrm{E},~33^{\circ}32'00''\mathrm{S}$);  and one in Chile,  the 1.3 m SMARTS telescope at the CTIO.

    Part of the ground-based follow-up data was taken using two 40 cm and two 60 cm telescopes from the Skynet network, PROMPT-5 and PROMPT-8 located at the CTIO, PROMPT-MO-1 located at the Meckering Observatory ($31^{\circ}38'17''\mathrm{S},~116^{\circ}59'20''\mathrm{E}$), and R-COP at the Perth Observatory ($32^{\circ}0'29''\mathrm{S},~116^{\circ}8'6''\mathrm{E}$).
    The object was also observed with the 60 cm Telescope Live CHI-1 telescope at El Sauce Observatory (Chile, $30^{\circ}28'21''\mathrm{S}, 70^{\circ}45'47''\mathrm{W}$) and the 45 cm Telescope Live AUS-1 instrument at the  Warrumbungle Observatory (Australia, $31^{\circ}16'35''\mathrm{S}, 149^{\circ}11'35''\mathrm{W}$), using Sloan $r'$, $g'$, and Johnson $V$ filters. In addition,  the Robotic Eye Mount (REM, ESO La Silla, $29^{\circ}15'\mathrm{S}$ and $70^{\circ}44\mathrm{W}$) telescope was used.
    
    All the data from the above-mentioned telescopes, except ROAD and $\mu$FUN, were standardised using the Cambridge Photometric Calibration Server\footnote{http://gsaweb.ast.cam.ac.uk/followup/} \citep{Zielinski2019,Zielinski2020},  a tool for coordinating observations from multiple sites and standardising the photometry, which is designed to help process and store the science-ready data of follow-up observations.
    
    \subsection{OGLE data}
    
\begin{figure}[t]
        \centering
        \includegraphics[width=0.9\columnwidth]{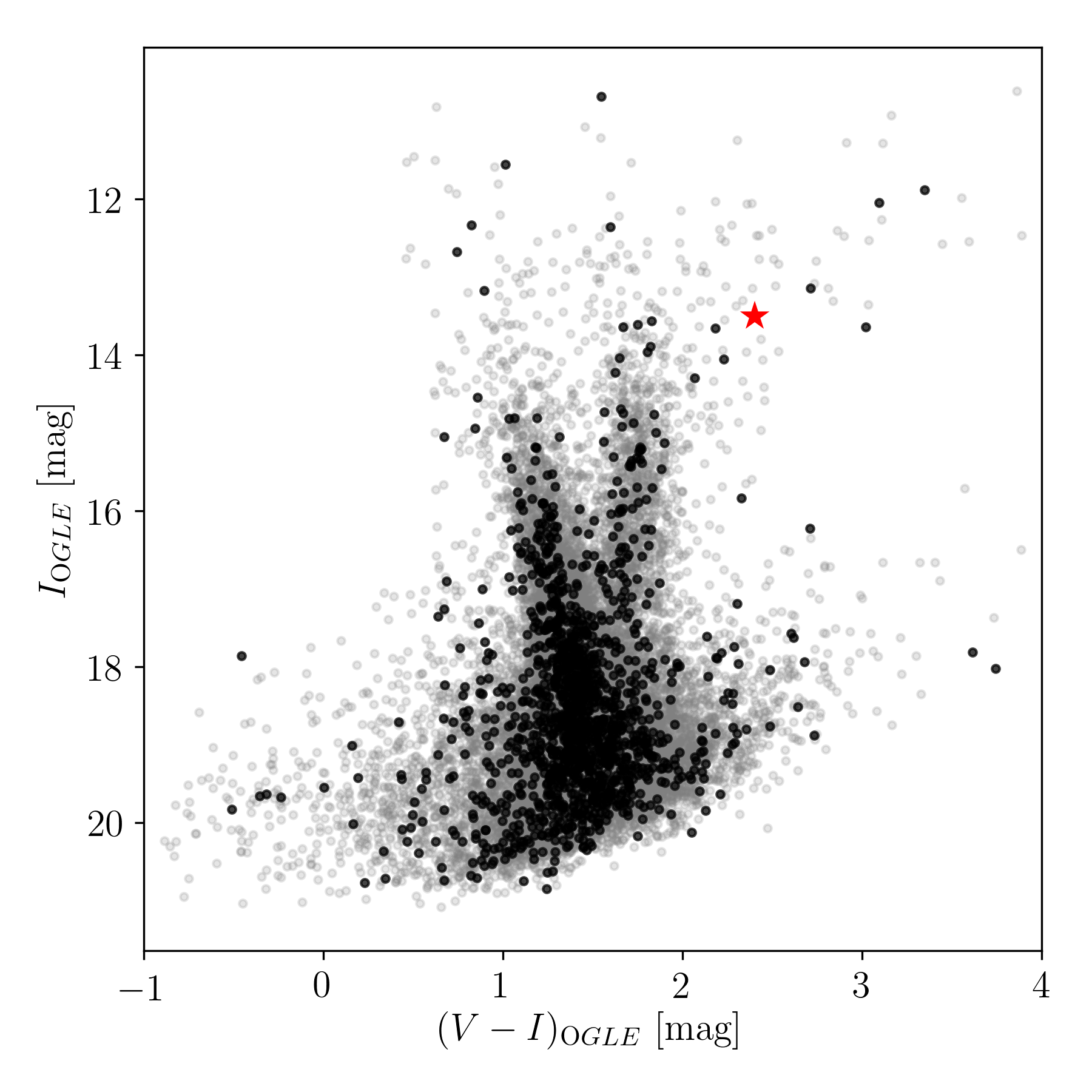}
        \caption{Colour-magnitude diagram of the field stars for Gaia19bld event. The grey dots represent stars from the whole OGLE field GD1298.15, while the stars closest to the location of the event (within the 5x5 arcmin box around the target) are highlighted in black. The position of the source is given by  the red star.}
        \label{cmd}
    \end{figure}

    The Gaia19bld event lies in the OGLE-IV disc field GD1298.15, and was already reported by \cite{Mroz2020} in the comprehensive statistical study of 630 microlensing events found by the OGLE project in the Galactic disc.
    OGLE provides a fairly complete light curve, because the cadence for this field was increased for the duration of the event from one per week to one per day. The OGLE project is currently in its fourth phase \citep{Udalski2015}; it uses a large mosaic camera with a field of view of 1.4 square degrees mounted on the 1.3m Warsaw University Telescope located at the Las Campanas Observatory, Chile ($29^{\circ}0'57''\mathrm{S},~70^{\circ}41'31''\mathrm{W}$). The OGLE project began in 1992 as a survey searching for microlensing events in the Galactic bulge and the Magellanic Clouds \citep{Udalski1992}, and in its fourth phase it also covered most of the southern Galactic disc. 
    Field GD1298.15 has been regularly monitored since March 2013. Figure \ref{fig_ogle_zoom} shows the neighbourhood of the Gaia19bld source star on the OGLE-IV reference image, while Figure \ref{cmd} presents the colour-magnitude diagram constructed using a calibrated OGLE-IV map for this field. 

    \subsection{Spitzer data}
    
    The Spitzer Space Telescope has been extensively used for the follow-up of microlensing events in recent years, which has resulted in multiple discoveries, especially in the field of extrasolar planets (e.g. \citealt{Udalski2015Spitzer, CalchiNovati2015, Zhu2017}). It is an important tool for this type of transient because it often allows  the microlensing parallax vector to be measured.
    
    Spitzer data were acquired, by request from the Director's office, to divert
    a small fraction of the observing time that had been allocated
    for Galactic bulge microlensing events during the 2019 season.
    Hence, the Spitzer observations began on 6 July,  at the
    beginning of this bulge microlensing programme, which was set
    by the observability of the bulge.  It was observed for 28
    days at a cadence of one per day (i.e. until the Spitzer observing
    window closed).

    Because the event was at maximum magnification on 16 July UT, conveniently for the microlensing parallax estimations, the Spitzer data cover the peak very well, so the shift between times of the event maximum $t_{0_{fup}}$ from the ground and $t_{0_{Spitzer}}$ from the satellite is easily measurable and yields $\Delta t_0 = t_{0_{Spitzer}} - t_{0_{fup}} = 2.8~\mathrm{days}$. For the reduction of the data we followed the standard procedure described in \cite{SpitzerReduction2015}.
    
    \subsection{Spectroscopic follow-up}
    
    In order to determine the atmospheric parameters as well as the distance to the source object, we  gathered several high-resolution spectra at various phases of brightness amplification. Two spectra were obtained  using the VLT/X-Shooter instrument (within ESO DDT proposal No. 2103.D-5046) around the peak of the light curve at July 29 and close to the baseline at November 28, 2019. In addition, two LCO/NRES spectra were obtained, both close to the peak on July 15 and 19, 2019 (proposal No. LCO2019B-014).
    
    Stellar parameters were extracted in two ways:  (i) by synthesising theoretical spectra and fitting particular line regions to the observed spectrum and (ii) by fitting the continuum of template spectra to the observational ones for specific effective temperature $T_{\rm eff}$, $\log{g}$, and metallicity. This method has allowed us to precisely determine the line-of-sight extinction $A_{V}$ and distance to the source. The parameters obtained based on this spectroscopic analysis are presented in Table \ref{spectral_params}.
    \begin{table}
    \centering
        \caption{Parameters of the source in the Gaia19bld event determined from spectroscopy. See B21 for more details.}
    \begin{tabular}{cc}
        \hline \hline
        \noalign{\smallskip}
        Parameter & Template matching \\
        \noalign{\smallskip}
        \hline   
        \noalign{\smallskip}
        $T_{\rm eff}$ [K] & $4097^{+32}_{-29}$\\ \\
        $\log{g}$ [dex] & $1.48^{+0.15}_{-0.16}$ \\ \\
        $[\rm M/H]$ & $0.295^{+0.053}_{-0.062}$ \\ \\
        $A_{V}$ [mag] & $2.322^{+0.075}_{-0.072}$ \\ \\
        $D_{\rm S}$ [kpc] & $8.4^{+0.8}_{-1.5}$ \\
        \noalign{\smallskip}
    \hline    
    \hline    
        
    \end{tabular}
    \label{spectral_params}
    \end{table}    
    The details of the spectroscopic analysis of the Gaia19bld event and the discussion of the results are presented in a complementary study B21. In this work we use spectroscopic parameters derived in  model A (see their Table 1), as it remains closest to the data and agrees best with $\theta_{\rm E}$ derived in C21. Using  the remaining models (B and C)  does not change the final results of this paper.
    
    \subsection{Systematic errors}
    
    All ground-based data collected near the peak of the event show systematic variations in the residuals from the best-fit model at a 1-3\% level (right panel of Figure \ref{lc}). These deviations are observatory-dependent, which indicates that they are likely due to low-level systematic errors in the photometric data rather than some unmodelled physical effects (such as star spots on the source surface).
The peak of the event coincided with the full moon and,  although the Moon was located about 100 deg away,  the elevated background may have introduced low-level systematics. To check this hypothesis, we investigated the light curves of a few nearby non-variable stars of similar brightness in the LCO data. We found that between HJD=2458675 and 2458688 (when the Moon phase was larger than 60\%) their rms scatter was twice as large as outside this period. To take this into account, the photometric error bars were increased by a factor of 2 for all ground-based data points collected during the above-mentioned period.
   
    \section{Model}
    
    \subsection{Data preparation}
    To optimise the modelling process we only use the most consistent data sets in our sample. We naturally exploit survey data sets (OGLE and Gaia), which provide both accurate light curves and archival photometric baselines. The Las Cumbres observations cover the light curve throughout the whole duration of the event and is the most uniform data set in the collected photometric follow-up. While it introduces some systematics, its consistency makes it one of our "core" data sets. Even though the Las Cumbres network consists of nearly identical telescopes, with the same instruments and filters, their characteristics can differ from site to site, and even for different telescopes within the same site. This difference can be important especially close to the peak, where finite source effects and limb darkening become prominent. To obtain the most accurate results during the modelling, we separate the Las Cumbres data into 16 (eight I-bands and eight V-bands) separate sets and treat every distinct telescope-filter pair as an independent data set. Two sets from the CTIO site ($I$-band and $V$-band from the lsc1m009 camera) are discarded due to their small number of data points.
    We also utilise all the data gathered by $\mu$FUN network -- they do not cover baseline of the event, but provide a very dense sampling during the peak of the event. We bin KKO data into 0.5 hr bins due to large amount of data points collected. 
    
    Data taken with the ROAD observatory show significant scatter, especially for the fainter parts of the light curve.  Nonetheless, they cover the whole duration of the event and are very homogeneous, and are thus used in the modelling process. Due to the large number of data points we use 0.5 hr bins for the ROAD photometry. We also include observations from the Spitzer satellite in our analysis. All the data used in the modelling process are listed in Table \ref{tab_model_obs}. The remaining data sets either did not provide the homogeneous coverage or introduced significant systematic errors and scatter compared to the utilised sets.
    
\begin{figure*}[!t]
        \centering
        \includegraphics[width=2.1\columnwidth]{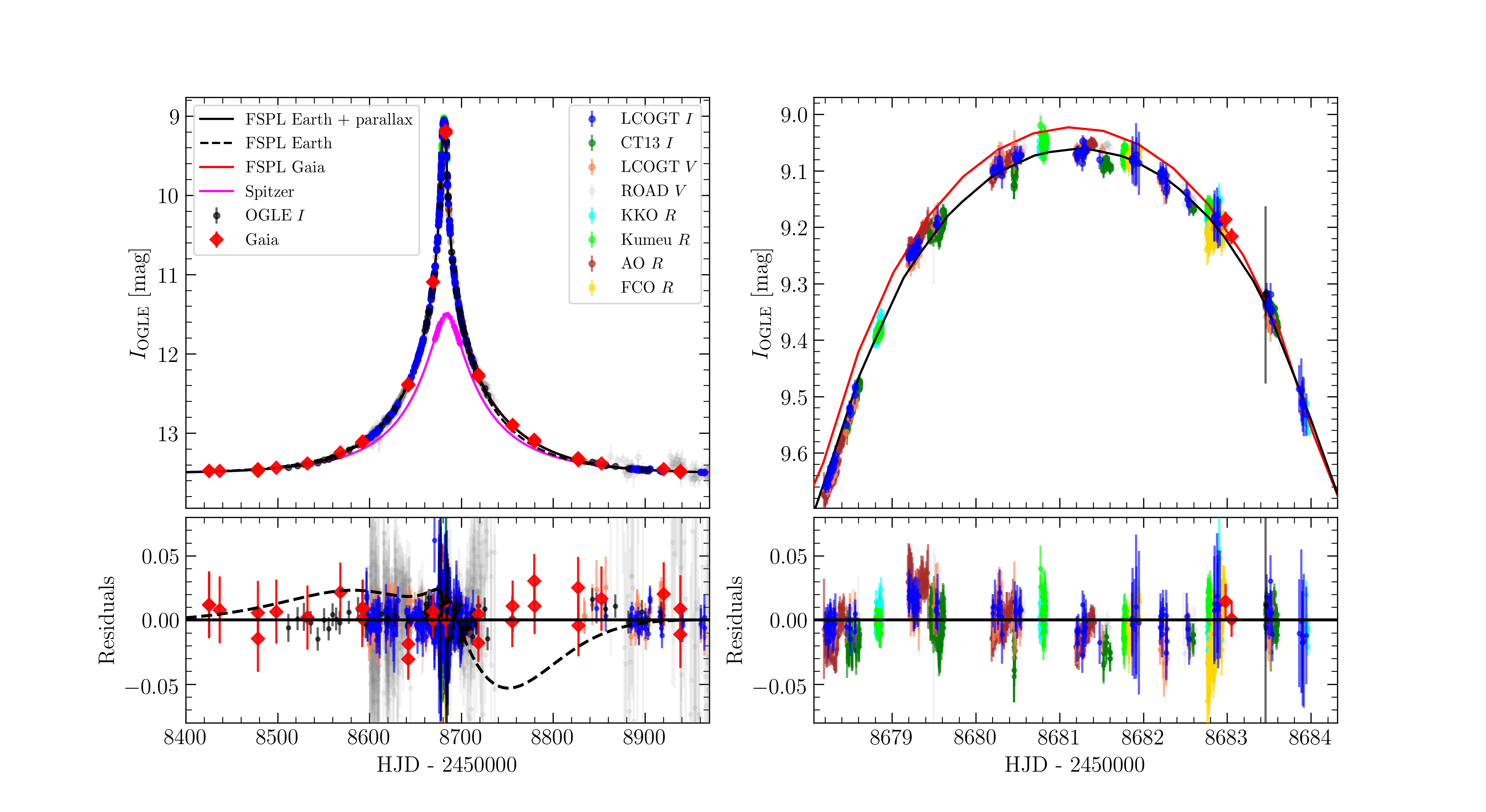}
        \caption{Ground-based and Gaia data sets used for the modelling, photometrically aligned to the $I_{\rm OGLE}$ magnitude. \textit {Left:} Light curve observed from Earth and Gaia, along with finite source models, with (black solid line) and without (black dashed line) parallax. As the Gaia satellite is located close to Earth, the observed magnification is very similar in both cases. \textit{Right:}  Peak of the event with finite source effects clearly visible, along with residuals from the model. The shape of the light curve as seen from the position of Gaia is slightly different. The importance of the intensive ground-based follow-up campaign is highlighted here.   The survey data (OGLE and Gaia) were sufficient to estimate standard microlensing parameters, while the high-cadence observations were vital to cover the peak and constrain the size of the source.}
        \label{lc}
    \end{figure*}

    We apply quality cuts on air mass, photometric error and seeing to the Las Cumbres data, mostly to eliminate images taken during the bad weather conditions and to obtain the highest quality sample without significant loss of coverage.  After the filtering, approximately 35\% of the Las Cumbres data were discarded.
    All the error bars were rescaled using the formula $\sigma_{i,\rm new} = \sqrt{(\gamma \sigma_{i})^2 + \epsilon^2}$ so that $\chi^2/dof \approx 1$ for the best model. For each data set we assumed a fixed value of $\epsilon = 0.001~{\rm mag}$ and then fit the $\gamma$ coefficient. The photometric data along with the light curve are plotted in Figure \ref{lc}.
    
    \begin{table}
      \centering
      \caption[]{Photometric data sets used for the modelling (after binning), removal of outliers,  and filtering out images taken during the bad weather. The $\gamma$ parameter is an error-rescaling factor.}
      \label{tab_model_obs}

         \begin{tabular}{cccc}
            \hline
            \hline
            \noalign{\smallskip}
            Observatory & Filters & Data points & $\gamma$\\
            \noalign{\smallskip}
            \hline
            \noalign{\smallskip}
            Gaia & $G$ & 79 & 4.4 \\
            Spitzer & 3.6 $\mu$m & 34 & 4.3 \\
            OGLE & $I$ & 210 & 1.78\\
            \hline
            LCOGT SSO (cleaned): \\
            \hline    
            coj1m003 & $I,V$ & 129 & 3.5, 1.5 \\
            coj1m011 & $I,V$ & 215 & 3.5, 2.8\\
            \hline
            LCOGT CTIO (cleaned): \\
            \hline
            lsc1m004 & $I,V$ & 120& 1.3, 1.8 \\
            lsc1m005 & $I,V$ & 164 & 1.2, 1.5\\
            \hline
            LCOGT SAAO (cleaned): \\
            \hline
            cpt1m010 & $I,V$ & 111 & 1.3, 1.8\\
            cpt1m012 & $I,V$ & 168 & 1.2, 1.3\\
            cpt1m013 & $I,V$ & 195 & 1.2, 1.3\\
            ROAD (binned) & $V$ & 676 & 1.2 \\
            \hline
            $\mu$FUN:\\
            \hline
            Kumeu & $R$ & 234 & 0.9\\
            KKO (binned) & $R$ & 175 & 3.0\\
            AO & $R$ & 439 & 0.9\\
            FCO & - & 311 & 0.7\\
            CT13 & $I$ & 535 & 0.5\\            
            \hline
            \hline
         \end{tabular}

   \end{table}

    \begin{figure}
        \centering
        \includegraphics[width=\columnwidth]{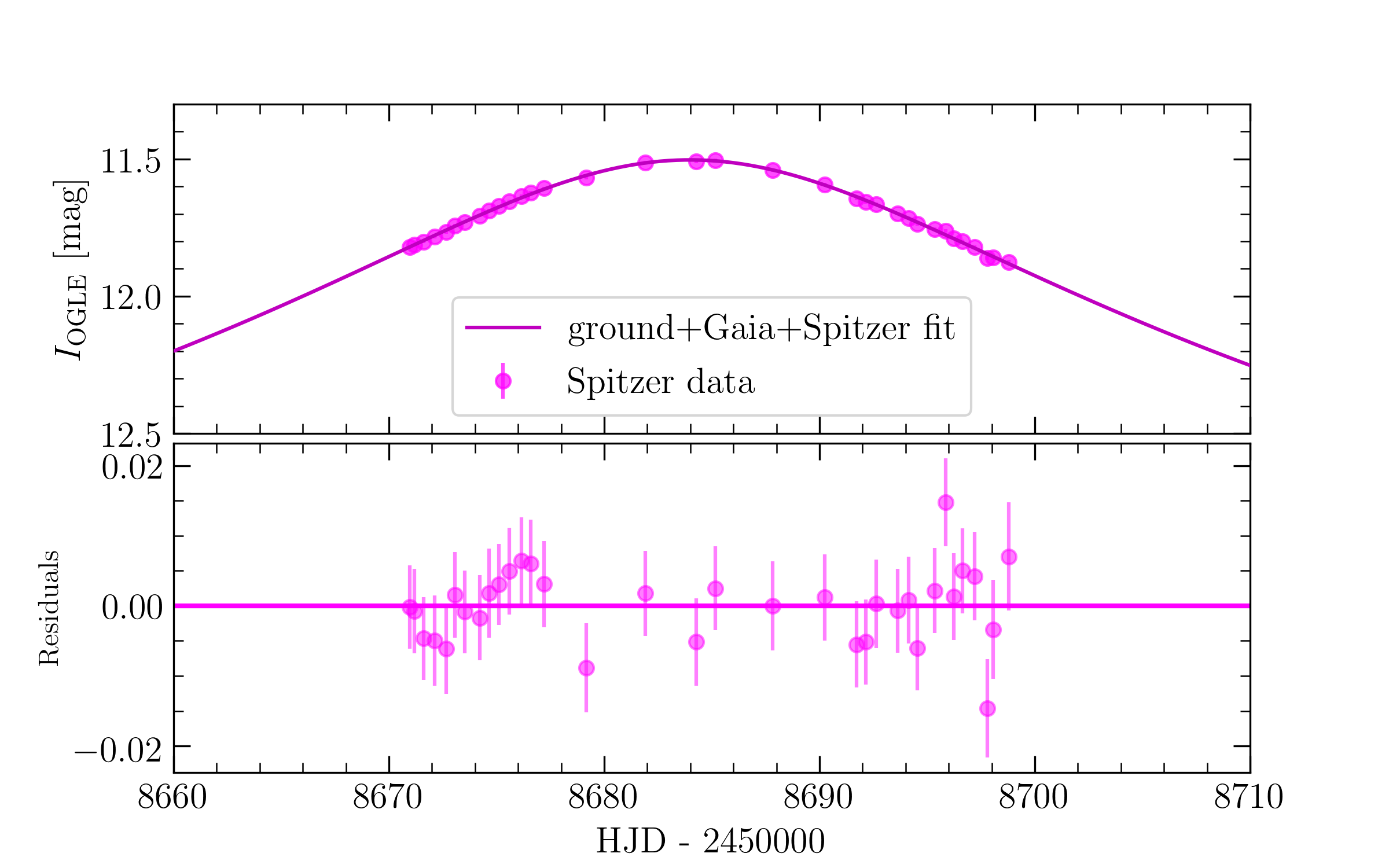}
        \caption{
        The light curve of the Gaia19bld event as seen by the Spitzer satellite, adjusted to the OGLE $I$ magnitude. While some systematics are visible in the first patch of the data, the effect is not critical and the microlensing (space) parallax measurement provided by Spitzer is robust.
        }
        \label{fig:Spitzer}
    \end{figure}
    
    \subsection{Single-lens model, finite source effect, and limb darkening}
    
For the modelling of this event we used the procedures provided in the pyLIMA package \citep{pyLIMA2017}. In such high-magnification events with upper-giant-branch sources, we can often expect finite source effects to be measurable.  The simplest  point-source, point-lens (PSPL) model does not reproduce the observed data. We employ the finite-source, point-lens (FSPL) model that includes the finite size of the source star \citep{Yoo2004}.
    Because the discs of stars as we observe them are not of uniform brightness, we implement a simple one-parameter model of linear limb darkening (LLD) so that the surface brightness $S$ is a function of the distance $r$ from the centre of the star \citep{Albrow1999_LD},
    \begin{equation}
        S_{\lambda}(r)=\bar{S}_{\lambda} \left(1-\Gamma_{\lambda} \left(1-\frac{3}{2}\sqrt{1-\left(\frac{r}{R_*}\right)^2} \right) \right),
    \end{equation}
    where $\bar{S}_{\lambda}$ is the mean surface brightness of the source, $R_*$ is its radius,  $\Gamma$ is a LLD coefficient, and the $\lambda$ index indicates dependence on the wavelength. Because $\Gamma$ is different for every filter we observed in, four separate parameters need to be introduced: $\Gamma_I$, $\Gamma_R$, $\Gamma_V$, and $\Gamma_G$ for $I$-band, $V$-band, $R$-band, and Gaia $G$-band observations, respectively.
    \begin{table}
    \centering
    
    \caption{Theoretical predictions for linear limb darkening coefficients in $I$, $R$, $V$ bands \citep{Claret2000} and Gaia $G$ band \citep{Claret2019} that were used during the modelling. In parentheses we quote a different parametrisation of this effect, where $u=3\Gamma/(2+\Gamma)$.}
    \begin{tabular}{ccc}
        \hline
        \hline
        Filter &  $\Gamma$ ($u$)  \\
        \hline    
        $I$ &  0.59 (0.68) \\
        $R$ &  0.72 (0.79) \\
        $V$ & 0.80 (0.86) \\
        $G$ &  0.72 (0.79) \\
        \hline
        \hline
    \end{tabular}
    \label{tab_LD}
    \end{table}
    Initially, we tried a free-fit for the LLD coefficients, but during the search for the final solution we fixed them to the theoretical values from \cite{Claret2000} (see Table \ref{tab_LD}). It is important to note that there is a large discrepancy between the theoretical predictions and coefficients found by minimising $\chi^2$. The observed discrepancy may result from systematic errors present in the data, whether it has a physical or instrumental origin. In either case, it is safer to fix LLD coefficients on theoretical values for two reasons. First of all, these coefficients depend on the temperature, surface gravity and metallicity of the source star, and all these are well known from spectroscopy. Secondly, allowing LLD coefficients to change during the fit may lead to distortions in other parameters. The tension between the microlensing parallax measurement from the ground-based data (annual parallax) and Spitzer (space parallax) is smaller when using the theoretical LLD values from \cite{Claret2000}.
    
    Finally, because there is a clear asymmetry in the ground-based light curve, indicating that the microlensing parallax signal is present in the data (see residuals in the left panel of Figure \ref{lc}), we include this effect in the model. Thus, our FSPL model has three standard microlensing parameters, $t_0$, $t_{\rm E}$, and $u_0$; two microlensing parallax vector components, $\pi_{\rm EN}$, $\pi_{\rm EE}$; and the $\rho$ parameter for the size of the source (see Formula \ref{eq_rho}). Additionally, the source flux $F_{\rm S}$ and blend flux $F_{\rm bl}$ are calculated separately for every distinct filter-telescope pair.
    
    \subsection{Microlensing parallax analysis}
   
    Measuring the microlensing parallax vector $\pi_{\rm E}$ is crucial in the context of deriving the mass of the lens (see Formula \ref{eq_mass}). For Gaia19bld we have three potential ways of determining the parallax  as  observed from multiple locations, from Earth, from the Gaia satellite in the ${\rm L}_2$ region,  and from the Spitzer satellite orbiting the Sun at approximately 1 au from  Earth. The ${\rm L}_2$ point is relatively close ($\approx1.5\times10^6~\mathrm{km}$ from Earth), so the expected signal due to the space parallax between Gaia and ground observations was very low. On the other hand, from Spitzer's perspective the projected lens-source separation was very different than from Earth, and thus a robust space parallax measurement was expected. As the separation  observed from the satellite was much larger, the observed magnification was smaller, and the finite source effect was not detectable from Spitzer (see Figures \ref{lc} and \ref{fig:Spitzer}). While the model light curve as seen from Gaia is slightly different than from Earth, it is not constrained by any data since the observations were collected only for epochs where the difference is insignificant compared to the data precision (see the right panel of Figure \ref{lc}). The microlensing parallax measurement have to rely on the two remaining effects:  annual parallax and space parallax from the Spitzer satellite. We are able to make two essentially independent measurements
    of the microlens parallax vector $\pi_{\rm E}$ using the
    ground-only and Spitzer-``only'' data sets.  We describe these
    in turn below.  We  ultimately show that there is some tension
    between these determinations, and we  discuss the resolution
    of these tensions.
    
    For the ground-only determination, we will eventually consider
    different subsets of the data.  However, we proceed in the same way for each.  We first find initial solutions using a differential evolution 
    genetic algorithm. Because the event lies far from the ecliptic plane
    ($\beta=-54^\circ$) and the light curve coverage is very complete, 
    it is reasonable to expect that
    the microlensing parallax parameters will be well determined.
    Next we evaluate the error contours using the Markov
    chain Monte Carlo
  method (MCMC; \citealt{EMCEE}), starting near the initial solution
    with the best $\chi^2$.  There is a well-established degeneracy 
    $u_0\leftrightarrow -u_0$ \citep{Smith2003}, which
    we investigate as a matter of due diligence, finding very similar
    values of $\chi^2$.  However, for the first time for any microlensing
    event, there is an independent measurement of the sign of $u_0$ from
    VLTI (C21).  That is, by measuring the change in the
    orientation of the two images at two successive epochs, they
    were able to show that $u_0<0$; in the case of the Kojima event, Dong et al. (2019) were not able to make such a determination because they had
    only a single epoch. Therefore, we only consider the $u_0<0$ solution. 
    
    \begin{figure}
        \centering
        \includegraphics[width=1\columnwidth]{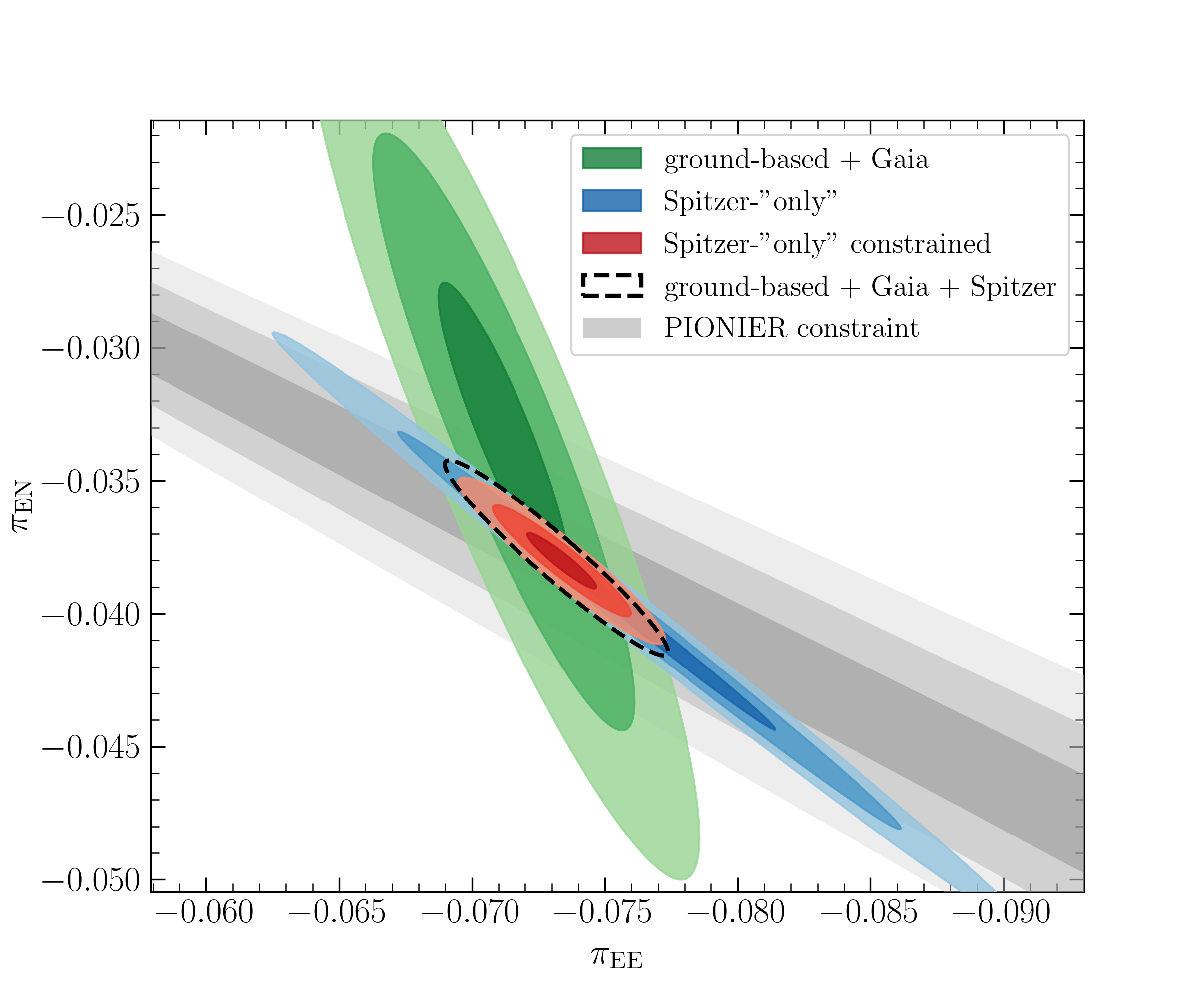}
        \caption{Limits of 1, 2, and 3$\sigma$  for microlensing parallax components for the $u_0 < 0$ solution derived using different methods and data sets. Here we present ground-based+Gaia contour (green, large ellipse), Spitzer-``only'' solution with free blending (blue elongated ellipse), and Spitzer-``only'' solution with a constraint on the blend flux (small red ellipse in the middle).  The black dashed ellipse shows the microlensing parallax components distribution obtained with a joint fit to ground-based, Gaia, and Spitzer data. It is almost identical to the solution obtained using the  Spitzer-``only'' constrained method. We also include 1, 2, and 3$\sigma$ PIONIER constraints derived in C21 (grey bands).}
        \label{Spitzer_only}
    \end{figure}
    
    \begin{table*}[h]
      \centering
      \caption[]{Solutions for Gaia19bld microlensing event, derived using different data sets. The difference $\Delta\chi^2 ~ 50$ between positive and negative $u_0$ scenarios points towards the latter. In addition, the interferometric measurements from C21 unambiguously determine the $u_0$ parameter to be negative, thus we adopt this solution as the real one.
      }
         \begin{tabular}{ccccc}
            \hline
            \hline
         
            \noalign{\smallskip}
            \multirow{2}{*}{Parameter}&ground-based+Gaia&ground-based+Gaia&ground-based+Gaia+Spitzer& ground-based+Gaia+Spitzer \\ &$u_0 > 0$ & $u_0 < 0$ & $u_0 > 0$ & $u_0 < 0$ \\
            \noalign{\smallskip}
            \hline
            \noalign{\smallskip}
            $t_{0, \rm par}$ [days] & 2458680.0 & 2458680.0 & 2458680.0 &  2458680.0\\   
            $t_0$ [days] & $2458681.210 ~\pm~ 0.001$ & $2458681.220~\pm~0.001$ & $2458681.208 ~\pm~ 0.001$ & $2458681.219~\pm~0.001$\\
            $u_0$ & $0.0193~\pm~ 0.0001$ & $-0.0193~\pm~ 0.0001$ & $0.0190~\pm~ 0.0001$ & $-0.0192~\pm~ 0.0001$\\
            $t_\mathrm{E}$ [days] & $106.17~\pm~ 0.55$ & $106.76~\pm~ 0.57$ & $107.60~\pm~ 0.46$ & $107.06~\pm~ 0.50$\\
            $\pi_{\mathrm{EN}}$  & $-0.0303~\pm~ 0.0056$ & $-0.0334~\pm ~0.0056$ & $-0.0415~\pm~ 0.0010$ & $-0.0378~\pm ~0.0012$\\
            $\pi_{\mathrm{EE}}$ & $-0.0684~\pm~ 0.0024$ & $-0.0713~\pm~ 0.0024$ & $-0.0757\pm~ 0.0012$ & $-0.0731~\pm~ 0.0014$\\
            $\rho$ & $0.03219~\pm ~0.00018$ & $0.03211~\pm ~0.00018$ & $0.03171~\pm ~0.00015$ & $0.03202~\pm ~0.00016$\\
            $\chi^2$ & 3068 & 3058 & 3134 & 3082 \\
            \hline
            \hline
         
         \end{tabular}
         \label{tab_sol}    
     \end{table*}
     
         \begin{table*}[h]
      \centering
      \caption[]{Parameter values derived using the Spitzer-``only'' method, which we employ here
      following \citep{Gould2020}.
     The results obtained with this approach and global ground+Gaia+Spitzer fit are almost identical. We do not list constrained Spitzer-``only'' for positive $u_0$ here because it did not converge to any solution, confirming the $u_0<0$ scenario.}
         \begin{tabular}{cccc}
            \hline
            \hline
         
            \noalign{\smallskip}
            \multirow{2}{*}{Parameter}&Spitzer-``only''&Spitzer-``only''& Spitzer-``only'' constrained \\ &$u_0 > 0$ & $u_0 < 0$ & $u_0 < 0$ \\
            \noalign{\smallskip}
            \hline
            \noalign{\smallskip}
            $t_{0, \rm par}$ [days] & 2458680.0 & 2458680.0 &  2458680.0\\   
            $t_0$ [days] & $2458681.210 ~\pm~ 0.001$ & $2458681.220~\pm~0.001$  & $2458681.220~\pm~0.001$\\
            $u_0$ & $0.0194~\pm~ 0.0001$ & $-0.0193~\pm~ 0.0001$ & $-0.0193~\pm~ 0.0001$\\
            $t_\mathrm{E}$ [days] & $106.0~\pm~ 0.7$ & $106.80~\pm~ 0.41$ & $106.77~\pm~ 0.42$\\
            $\pi_{\mathrm{EN}}$  & $-0.0557~\pm~ 0.003$ & $-0.0404~\pm ~0.0035$ & $-0.0381~\pm ~0.0013$\\
            $\pi_{\mathrm{EE}}$ & $-0.0944~\pm~ 0.004$ & $-0.0764~\pm~ 0.0044$ & $-0.0735~\pm~ 0.0016$\\
            $\rho$ & $0.03219~\pm ~0.00012$ & $0.03211~\pm ~0.00012$ & $0.03211~\pm ~0.00013$\\
            $\chi^2$ & 23 & 24 & 23 \\
            \hline
            \hline
         
         \end{tabular}
         \label{tab_sol_S}    
     \end{table*}       
    
    For the Spitzer-``only'' determination, we follow the method developed
    and described in detail
    by \cite{Gould2020} for KMT-2018-BLG-0029 and subsequently applied by \cite{Hirao2020} and \cite{Zang2020} to OGLE-2017-BLG-0406 and OGLE-2018-BLG-799, respectively.  We first fix the \cite{Paczynski1986} parameters $(t_0,u_0,t_{\rm E})$ at their ground-based values.
    Next we evaluate the Spitzer source flux $F_{\mathrm{S},Spitzer}=473\pm 19$ (in instrumental units)
    by combining
    a $VIL$ colour-colour relation (derived by matching Spitzer and OGLE
    field-star photometry) with the measured OGLE source flux, $F_{\rm S,OGLE}$. Finally,
    we fit only the Spitzer data.  In the initial version of this
    fit, there are three free parameters 
    $(\pi_{\rm EN},\pi_{\rm EE},F_{\mathrm{bl},Spitzer})$, and also one highly
    constrained parameter (from the $VIL$) relation:  $F_{\mathrm{S},Spitzer}$.
    This fit shows that that $F_{\mathrm{bl},Spitzer}=-40\pm 170$ is consistent with 
    zero, but with a relatively large error bar.  We therefore conduct a 
    second fit where we constrain $F_{\mathrm{bl},Spitzer}=0\pm 100$, which
    is very conservative based on the historical experience
    with the level of spurious negative blending generated
    by the photometry routine, and also with possible real positive blending
    given the constraints from the optical light curve.
    In principle, we should also consider the known
    four-fold degeneracy (\citealt{Refsdal964}, see also Figure 1 of \citealt{Gould1994}).  However, two of these degenerate solutions have $u_0>0$
    and so are eliminated by the constraint from C21 mentioned above. The third is strongly excluded by ground-based data.

    The Spitzer-``only'' method is a means to isolate the parallax information coming from the Spitzer data alone. Because it is \textit{not} a joint fit, the resulting parallax measurement will not be as affected by systematic errors in the ground-based data, whereas it might be in a global fit to all data sets. In previous applications of this method (\citealt{Gould2020, Hirao2020, Zang2020}) it has been important for isolating and evaluating potential effects due to systematic errors in the Spitzer photometry. In these three cases, the Spitzer detection was weak and the photometry only covered the falling wing of the light curve. Hence, in these cases, systematic errors at the level of 1-2 instrumental flux units were potentially significant. By contrast, Gaia19bld is strongly detected by Spitzer, and the Spitzer photometry covers the peak of the light curve. Hence, the role of systematics in the Spitzer data for this event should be minimal. However, because we have parallax information from several sources (ground-based photometry, Spitzer, PIONIER), we used the Spitzer-``only'' method in order to understand the constraints on the parallax contributed by Spitzer alone. We also performed a joint, global fit to all the photometric datasets (ground-based + Gaia + Spitzer), and find the results are nearly identical to the Spitzer-``only'' constraints.

In Tables \ref{tab_sol} and \ref{tab_sol_S} we illustrate differences in the approaches to modelling described above. Constraints on the microlensing parallax vector $\pi_{\rm E}$ are shown in Figure \ref{Spitzer_only}. The grey bands represent the constraint on the direction on the microlensing parallax vector given by PIONIER measurements (C21). While there is a slight tension between the annual and space parallax measurements, all three methods are consistent, and the constraint from interferometry agrees very well with the Spitzer result.
    
    \section{Physical parameters of the source and lens}
    All the information about the size of the source star we can get from the light curve are contained in the $\rho$ parameter, which is the angular radius of the source star expressed in the Einstein radii units. To calculate $\theta_{\rm E}$ we also need an estimate of the angular size of the source $\theta_*$ (see Formula \ref{eq_rho}). The usual procedure is to use empirical relations between the source colour and its angular size (e.g. \citealt{Boyajian2014}, \citealt{Adams2018}). In this case we have access to more robust measurement derived from spectroscopy (B21):
    \begin{equation}
    \theta_* = 24.16^{+0.39}_{-0.40}~\mu \mathrm{as}
    .\end{equation}    
    This value is used in all the calculations. We note that angular size derived using the photometric method yields $\theta_* = 24.5~\pm 2.4~\mu \mathrm{as}$. While it is less accurate, it remains consistent with the spectroscopic derivation.
    
    Accurate source distance determination for the Gaia19bld event was possible thanks to the spectroscopic follow-up, which is discussed in detail in B21. Among the multiple methods described there, we chose the one for which the absolute magnitude of the source is derived using the isochrone with fixed age of 1 Gyr, knowing the source temperature, metallicity, $\log{g}$, and extinction from spectrum fitting (see Table \ref{spectral_params}). This method yielded
    \begin{equation}
        D_{\rm S} = 8.4^{+0.8}_{-1.5} ~\mathrm{kpc}
    .\end{equation} 
    The parallax signal detected by Gaia is weak and thus the distance determined by the satellite is very uncertain (see Table \ref{tab_GDR2}).
    In \cite{CBJ2020}, where EDR3 measurements are supported by a simple Galactic model used as a prior to construct a posterior distribution of distances, the reported distance to Gaia19bld is $D_{\rm S, BJ} = 8.2^{+1.6}_{-1.1}~\rm kpc$. In principle, the astrometric parameters obtained by Gaia for a microlensing event can be altered by the presence of (at least) two light sources along the line of sight. Because the blending for this event is almost non-existent, we assume that the astrometric solution from EDR3 is for the microlensed source. The Gaia-based distance estimate agrees with the spectroscopic one, but we adopt the spectroscopic determination because it is  more robust.
    \begin{table}
        \centering
        \caption{Astrometric parameters measured by the Gaia satellite and reported in the EDR3 catalogue for the source along the line of sight. The three columns show the annual parallax and two proper motion components,  in the direction of increasing right ascension and that of increasing declination.}
        \begin{tabular}{c|c|c}
        \hline
        \hline
        $\pi~\rm [mas]$ & $\mu_{\alpha*}~\rm [mas/yr]$ & $\mu_{\delta}~\rm [mas/yr]$ \\
        \hline
        $0.08\pm0.02$ & $-7.43\pm0.02$ & $0.09 \pm0.02$ \\
        \hline
        \hline
        
        \end{tabular}

        \label{tab_GDR2}
    \end{table}

    \begin{table*}
      \centering
      \caption[]{Physical parameters of the single lens in Gaia19bld computed for different modelling approaches.
      All three  agree very well, while the constrained Spitzer-``only'' method and global fit to ground+Gaia+Spitzer data yield almost identical results. The values of $M_{\rm L}$ and $D_{\rm L}$ were derived from the measured values of $\theta_{*}$ and $D_{\rm S}$ by B21.
      }
      \label{tab_physical}

         \begin{tabular}{cccc}
            \hline
            \hline
            \noalign{\smallskip}
            \multirow{2}{*}{Parameter} & ground-based & Spitzer-``only''&ground-based \\
            & +Gaia & constrained & +Gaia +Spitzer
            \\
            \hline
            \noalign{\smallskip}
            $\pi_{\mathrm{E}}$   & $0.0786^{+0.004}_{-0.005}$& $0.0828^{+0.0027}_{-0.0020}$&$0.0823^{+0.0018}_{-0.0018}$ \\ \\
            $D_{\rm L}~\mathrm{[kpc]}$  & $5.61^{+0.42}_{-0.72}$& $5.53^{+0.34}_{-0.54}$&$5.52^{+0.35}_{-0.64}$ \\ \\
            $M_{\rm L}~[M_{\odot}]$  & $1.18^{+0.07}_{-0.06}$& $1.12^{+0.03}_{-0.03}$&$1.13^{+0.03}_{-0.03}$ \\ \\
            $\mu_{\rm rel, hel, N}~\rm [mas/yr]$&$-1.39\pm0.16$&$-1.50\pm0.04$&$-1.49 \pm 0.04$\\ \\
            $\mu_{\rm rel, hel, E}~\rm [mas/yr]$&$-2.39\pm0.16$&$-2.37 \pm 0.05$&$-2.36 \pm 0.05$\\\\
            $\mu_{\mathrm{ L, hel}, l}~\rm [mas/yr]$&$-9.74\pm0.16$&$-9.70\pm0.05$&$-9.69 \pm 0.05$\\ \\
            $\mu_{\mathrm{ L, hel}, b}~\rm [mas/yr]$&$-1.83\pm0.16$&$-1.93\pm0.05$&$-1.93 \pm 0.05$\\     
        \noalign{\smallskip}    
         \hline
         \hline
         \end{tabular}
             
     \end{table*}         

     Using the angular size of the source and second equation from Formula \ref{eq_rho}, we derive the angular size of the Einstein radius
     \begin{equation}
         \theta_{\mathrm{E}} = 0.755 \pm 0.013~ \mathrm{mas}
     ,\end{equation}
     which is in agreement with the value derived independently from VLTI interferometry (C21).
     Combining $\theta_{\mathrm{E}}$ with the parallax measurement $\pi_{\mathrm{E}}$, we can derive the mass of the lens;  using the second equation from Formula \ref{eq_mass}, we can derive its distance (see Table \ref{tab_physical}).
     
     The last physical characteristic of the source and lens that can be derived from the available parameters is their proper motion. As we measured $\theta_{\rm E}$ and $t_{\rm E}$, from the first equation in Formula \ref{eq_theta} we have $\mu_{\rm rel,geo} = 2.58^{+0.05}_{-0.05}~ \rm mas/yr$,          where the subscript `geo' is used to denote the geocentric frame adopted during the modelling. Because the direction of the relative proper motion vector is the same as the microlensing parallax vector (see Formula \ref{eq_piE}), it can be written as
     \begin{equation}
         \vec{\mu}_{\rm rel,geo}~(\rm N, E) = \mu_{\rm rel, geo}\frac{\vec{\pi}_{\rm E}}{\pi_{\rm E}} = (-1.18, -2.29)~\rm mas/yr
     .\end{equation}
     It is useful to transform it to the heliocentric frame so that it can be directly used for future adaptive optics observations:
     \begin{equation}
         \vec{\mu}_{\rm rel, hel} = \vec{\mu}_{\rm rel,geo} + \vec{v}_{\perp, \rm \oplus}\frac{\pi_{\rm rel}}{\rm au}
     .\end{equation}
     Knowing $\theta_{\rm E}$ and $\pi_{\rm E}$, we can calculate $\pi_{\rm rel} = 0.064 \pm 0.0064~ \rm mas$ (see Formula \ref{eq_piE}). The projected Earth velocity at the time $t_{\rm 0,par}$ is $\vec{v_{\perp, \rm \oplus}}~(\rm N,E) = (-24.1, -5.86)~\rm km/s$, and thus the relative heliocentric proper motion of the source and lens  $\vec{\mu}_{\rm rel, hel}~(\rm N,E) = (-1.49,-2.36)~\rm mas/yr$. For the source proper motion we adopt the value measured by Gaia, so $\vec{\mu}_{\rm s,  hel}~(\rm N,E) = (0.09, -7.43)~\rm mas/yr$. Because $\vec{\mu}_{\rm L} = \vec{\mu}_{\rm s} + \vec{\mu}_{\rm rel}$, we can calculate $\vec{\mu}_{\rm L, hel}~(\rm N, E) = (-1.40, -9.79)~\rm mas/yr$. After rotating this vector to the Galactic coordinates, we obtain $\vec{\mu}_{\rm L,  hel}~(l,b) = (-9.69, -1.93)~\rm mas/yr$. We then compare this result to the Besan\c{c}on Galactic model \citep{Besancon}. We simulate all stellar populations within a $0.2~ \rm deg^2$ field of view towards the event, lying at the lens distance. We find proper motions in this region of the Galaxy to be $\vec{\mu}_{D_L}~(l,b) = (-7.00, -0.35)\pm(1.7, 1.2)~\rm mas/yr$ for the thin-disc population. To further verify proper motions in this region of the Galaxy, we employ Gaia EDR3 catalogue and distance estimates from \cite{CBJ2020}, which yields $\vec{\mu}_{D_L}~(l,b) = (-6.5, -0.4)\pm(2.4, 1.8)~\rm mas/yr$. The measured lens proper motion lies well within 3$\sigma$ of both theoretical and EDR3-based distributions, which suggests that the lens does not have any significant motion with respect to the local rotation curve, and belongs to the disc population. The proper motions derived here are listed in Table \ref{tab_physical}

    \begin{figure}
        \centering
        \includegraphics[width=\columnwidth]{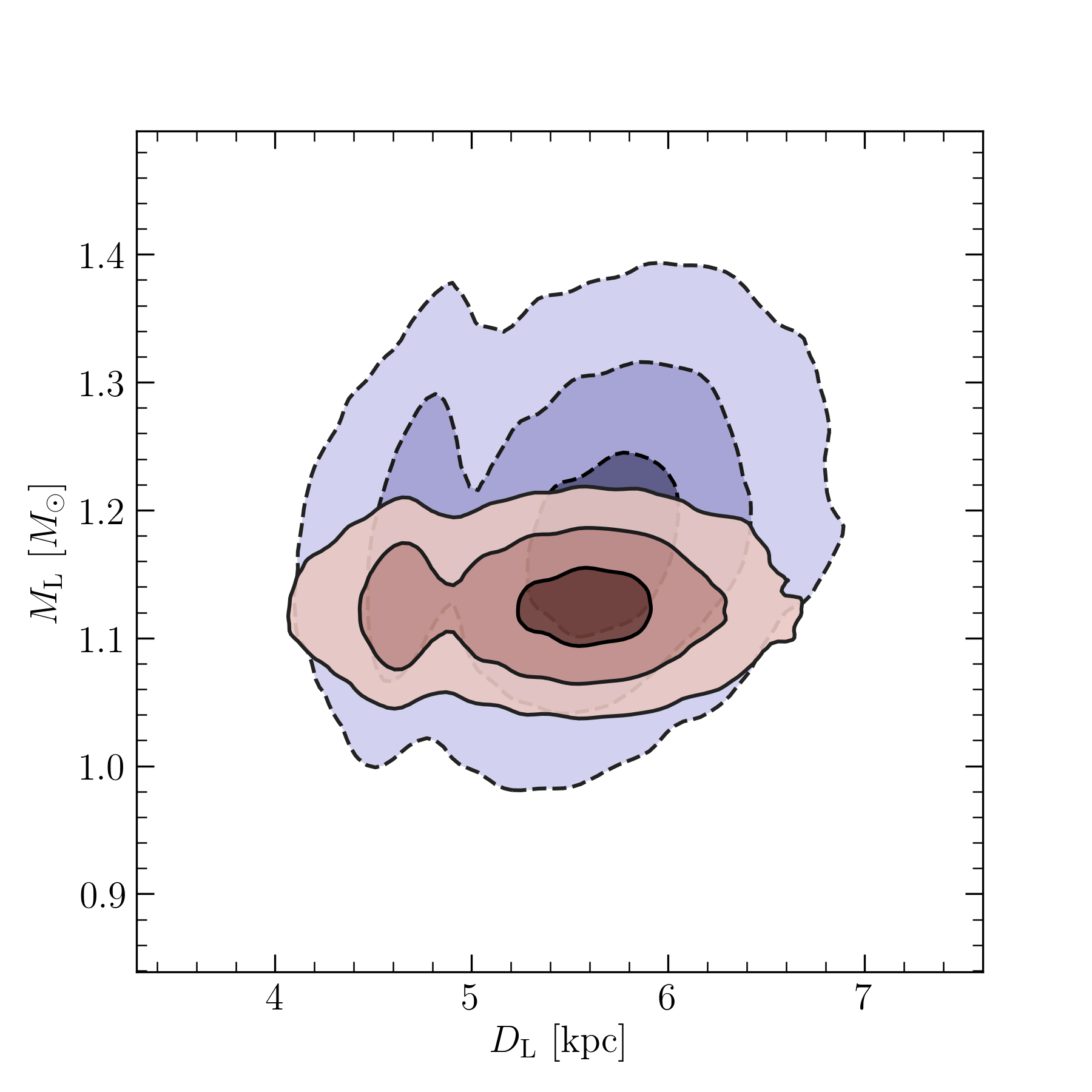}
        \caption{
        Posterior distribution of the distance to the lens $D_{\rm L}$ and its mass $M_{\rm L}$ for the ground+Gaia+Spitzer model (red solid contours) and the ground+Gaia model (blue dashed contours). The shaded areas show the  1, 2, and 3$\sigma$ contours for each of the solutions. As the $u_0$ parameter is known to be negative from C21, only $u_0 < 0$ solutions are shown here. The additional clump at $D_{\rm L}\approx 4.5$ kpc is caused by irregularities in source distance distribution calculated in B21.
        }
        \label{M_Dl}
    \end{figure}      

    \section{Discussion and conclusions}
    
    \subsection{Nature of the lens}
    
    The accurate photometric model and additional information about the source from spectroscopy, allowed us to determine the mass of the lens with good accuracy (see Table \ref{tab_physical}), but this is only a first step in the process of lens characterisation;  the ultimate goal is to reveal its actual nature and investigate whether the object is a regular star or a  rarer stellar remnant. Here we followed the strategy of examining the blend light after \cite{Wyrzykowski2016} to calculate probability that the lens is a stellar remnant. Knowing only the mass of the lens is not enough because a distant main sequence star can be as faint as a nearby white dwarf, neutron star, or a black hole. With the distance to the lens derived from eq. \ref{eq_mass} and the empirical mass-luminosity relation for main sequence stars (e.g. \citealt{Mamajek1}), this degeneracy can sometimes be broken because it might be possible to estimate the expected brightness of the lens $I_{\rm MS}$, as if it were a main sequence star. We also adopted the extinction value of $A_I=1.41^{+0.06}_{-0.03}~\mathrm{mag}$ determined from the spectra (B21). While this is the extinction to the source, we note that most of the dust resides between the Earth and the lens, and thus we use this value as the upper limit on extinction. 
    Figure \ref{M_Dl} shows the posterior distributions of lens mass and its distance. Using these two parameters, the  mass-luminosity relation and extinction $A_I$, we found the expected main sequence lens brightness to be $I_{\rm MS} = 18.71\pm0.12$ mag.

    In order to assess the amount of light from the blend accurately, we need a precise estimation of the baseline brightness of the event and the blending parameter from the microlensing model. 
    When estimating the blended light in bulge microlensing events,
    it is important to take account of the mottled background
    of unresolved stars because, if the source falls in a hole
    in this background, it will lead to an underestimate of the
    blended light, and thus to an overly strict limit on the lens light
    \citep{mb03037}.  \citet{Gould2020} gave a prescription for 
    estimating this effect that employs the \citet{holtzman98} luminosity function based on Hubble Space Telescope observations of Baade's Window. Previously, this effect had been estimated less precisely.  For example, the blended light given by \citet{mb13220}, which turns out to be the 90\% confidence limit based on the \citet{Gould2020} prescription,
    was subsequently shown to be significantly underestimated based on
    the adaptive optics observations of \citet{mb13220b}.  Nevertheless, despite the demonstrated importance of this issue for bulge microlensing, it is hardly relevant for lenses that, like Gaia19bld, lie far from the Galactic bulge because the surface density of unresolved stars is dramatically smaller.
    
    While using the whole multi-site data set is necessary for constraining the parallax and size of the source, to estimate the baseline brightness and blending parameter we analyse the OGLE data only. As already stated, the OGLE data covered the source years before the event and have a confirmed history of stable and precise photometry. From fitting the model to the OGLE data alone with constraint on $\rho$ parameter (the OGLE data did not cover the very peak of the event) and microlensing parallax vector $\pi_{\rm E}$, we obtain $$f_{s, \mathrm{OGLE}}=1.029\pm0.016, ~~~~~~~~ I_{0,\mathrm{OGLE}}=13.483\pm0.016. $$
    The value of the blending parameter is very close to unity, which indicates that there is essentially no extra light contributing to the baseline, and hence that  the blend is very faint or dark. Theoretically, $f_s > 1$ implies that the flux from the blend is negative, but it is most likely a consequence of the data reduction process and  not due to the physical properties of the lens. There is also a possibility that the blend is a light source unrelated with the event (being neither amplified nor the lens), but simply located within the seeing disc of the source star. However, because Gaia19bld is located in the Galactic disc region of relatively sparse stellar density, we assume that any blend light is to be attributed to the lens.

    To derive the upper limit on the brightness of the lens from blending in the microlensing model, we used values of the blending parameter $f_{s, \mathrm{OGLE}}$ and the baseline brightness $I_{0,\mathrm{OGLE}}$ $3\sigma$ away from the centre of their respective posterior distributions. From this we obtain $I_{\rm lens}>17.99$ mag, which is brighter than the theoretical brightness $I_{\rm MS}=18.71\pm0.12$ mag. It means that the microlensing model shows that there is enough light from the lens for it to be a main sequence star at the 3$\sigma$ level. Nonetheless, the overall blending solution suggests that the lens may be dark. Even so, this result is uncertain and at this point we cannot conclude on the nature of the lens.

More detailed observations, in particular in the ultraviolet part of the spectrum of the source, could potentially reveal an excess coming from a white dwarf lens. Imaging with Hubble Space Telescope (HST) might also reveal an additional object near the line of sight if the blend flux is not related to the lens. Nonetheless, it is rather unlikely as the event is located in the Galactic disc, where the probability of such a configuration is much lower than in the case of Bulge events. The derived parameters of the event can be used to accurately compute the expected location of the lens. If it is luminous (a faint main sequence star or a white dwarf), it will be possible to resolve it in a decade or so, using existing (adaptive optics infrastructure, HST) or future (e.g. ESO's ELT) instruments. Lack of detection of the lens could open up an exotic possibility that the lens was a low-mass black hole of primordial origin (e.g. \citealt{Carr2021, Carr2018, Garcia-Bellido2018}). 
    
    \subsection{Astrometric microlensing prospects}
    
\begin{figure}[h]
        \centering
        \includegraphics[width=1\columnwidth]{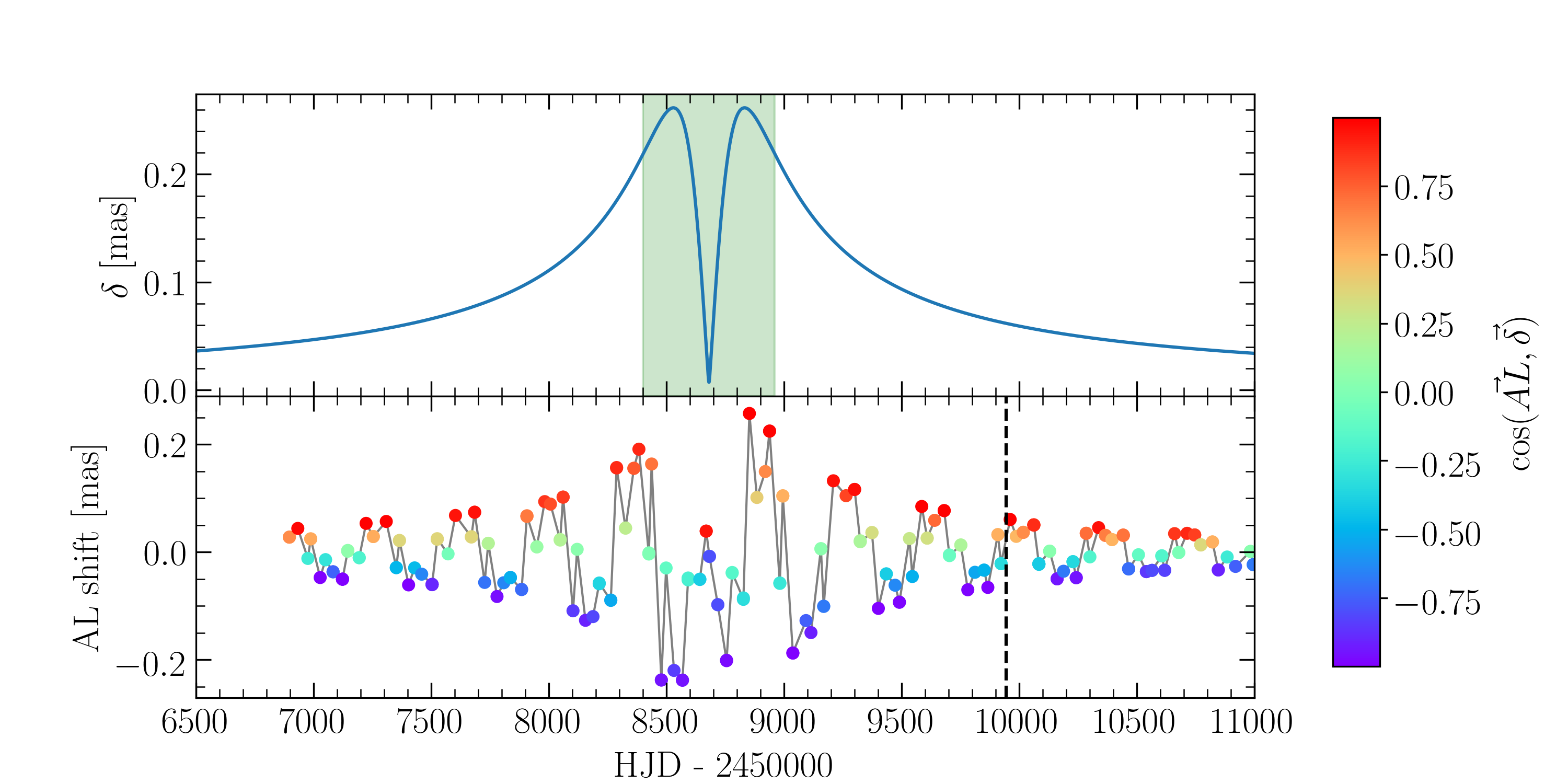}
        \caption{Simulations of the Gaia astrometric measurements. Top panel: Absolute shift due to the astrometric microlensing. The  green shaded region indicates the span of the photometric event. Bottom panel: Projection of the astrometric microlensing displacement  along the scan direction. There is no noise included in the simulation. The angle between the scanning direction and the centroid shift $\vec{\delta}$ due to microlensing is colour-coded (see colour bar to the right). The end of the extended mission (currently December 2022) is indicated by  the vertical dashed line.}

        \label{1D}
    \end{figure}

    Yet another way to measure the angular Einstein radii for some microlensing events will soon be possible with the Gaia astrometric data, which will be delivered with Data Release 4 (DR4) around 2023 or later. We simulated the Gaia astrometric time series for Gaia19bld based on the characteristics of the spacecraft and its data. Because in this case we know the value of $\theta_{\rm E}$, as well as the exact timings of the Gaia observations and the relative scanning directions for every epoch (Gaia Observation Forecast tool\footnote{https://gaia.esac.esa.int/gost/}), we project the astrometric displacements on the directions of scans to generate the measurements of Gaia, shown in Figure \ref{1D}. As shown in \cite{Lee_FS_astro2010}, the finite source effect is detectable in the astrometric microlensing signal. Nonetheless, it significantly affects the centroid trajectory mostly during the very peak of the event, while all but one of Gaia's observations were taken outside of this period. Because it does not impact the result, for simplicity we consider a point source model in our Gaia astrometric data simulations. We do not include any observational noise here. For the brightness similar to the baseline of Gaia19bld, the astrometric accuracy in the AL direction is expected to be around 0.1 mas \citep{Rybicki2018}, while the anomaly amplitude is around $0.25$ mas. Therefore, it may be possible to detect the astrometric microlensing signal in Gaia DR4 data and thus to estimate the $\theta_{\rm E}$ parameter for this event independently. 
    
    \section{Summary}

    In this work we have analysed the light curve of the microlensing event Gaia19bld and derived the mass of the lens, which appears to be a single object. This was made possible by measuring $\theta_\mathrm{E}$ during the central crossing of the lens in front of the source star disc (finite source effect) and by obtaining the microlensing parallax from ground- and space-based observations. We derived a lens mass of $M_{\rm L} =1.13 \pm 0.03 ~ M_{\odot}$ and distance $D_{\rm L}=5.52^{+0.35}_{-0.64} ~\mathrm{kpc}$. We analysed the blended light in order to assess the possibility that the lens is luminous. 
    The microlensing solution suggests that there is no blended light, and thus that the lens is dark, but we cannot give a definite answer about its nature as a distant main sequence star could be still acting as the lens. The dark lens scenario can be verified in high angular resolution imaging in about a decade's time when the lens and the source separate by about 30 mas; a luminous lens should be brighter than $\approx$18.7 mag.
    
    Gaia19bld is a spectacular showcase of the possibilities flowing from multi-faceted observations of microlensing events. Using two space satellites, ground-based survey telescopes, a follow-up network of smaller telescopes, high-resolution spectroscopy, and interferometric measurements, it was possible to accurately characterise the lens and the source, which would not have been possible with any of these channels individually. 
    Additionally, it may be possible to further verify these results in the future. With the Gaia astrometric time series data that will become available at the end of the mission, one could expect a measurable astrometric microlensing signal in this event. Even though the Einstein radius is not very large in this case, the brightness is extremely high, which is crucial for precise Gaia measurements.
    Although the methods applied for this event (finite source effects and interferometry) can only be used in very special cases, the future prospects for routine measurements of $\theta_\mathrm{E}$ via astrometric microlensing effect with Gaia and later with the Roman Space Telescope, look very promising, especially in the context of deriving mass distributions of invisible objects like neutron stars or black holes.
    
   \begin{acknowledgements}
The authors would like to thank the referee for their comments, which helped improving the quality of the paper. This work was supported from the Polish NCN grants: Preludium No. 2017/25/N/ST9/01253, Harmonia No. 2018/30/M/ST9/00311, MNiSW grant DIR/WK/2018/12, Daina No. 2017/27/L/ST9/03221, and by the Research Council of Lithuania, grant No. S-LL-19-2.
The OGLE project has received funding from the NCN grant MAESTRO 2014/14/A/ST9/00121 to AU.
We acknowledge the European Commission's H2020 OPTICON grant No. 730890.
YT acknowledges the support of DFG priority program SPP 1992 "Exploring the Diversity of Extrasolar Planets" (WA 1047/11-1).
EB and RS gratefully acknowledge support from NASA grant 80NSSC19K0291.
Work by AG was supported by JPL grant 1500811.
Work by JCY was supported by JPL grant 1571564.
SJF thanks Telescope Live for access to their telescope network.
NN acknowledges the support of Data Science Research Center, Chiang Mai University.
F.O.E. acknowledges the support from the FONDECYT grant nr. 1201223.
MK acknowledges the support from the NCN grant No. 2017/27/B/ST9/02727

\end{acknowledgements}

%
%

\bibliographystyle{aa} 
\bibliography{Gaia19bld_v7}

\begin{thebibliography}{96}
\expandafter\ifx\csname natexlab\endcsname\relax\def\natexlab#1{#1}\fi

\bibitem[{{Abbott} {et~al.}(2016){Abbott}, {Abbott}, {Abbott}, {Abernathy},
  {Acernese}, {Ackley}, {Adams}, {Adams}, {Addesso}, {Adhikari}, \&
  et~al.}]{Abbott2016}
{Abbott}, B.~P., {Abbott}, R., {Abbott}, T.~D., {et~al.} 2016, Physical Review
  Letters, 116, 061102

\bibitem[{{Abbott} {et~al.}(2017){Abbott}, {Abbott}, {Abbott}, {Acernese},
  {Ackley}, {Adams}, {Adams}, {Addesso}, {Adhikari}, {Adya}, \&
  et~al.}]{Abbott2017Multi}
{Abbott}, B.~P., {Abbott}, R., {Abbott}, T.~D., {et~al.} 2017, \apjl, 848, L12

\bibitem[{{Ackley} {et~al.}(2020){Ackley}, {Amati}, {Barbieri}, {Bauer},
  {Benetti}, {Bernardini}, {Bhirombhakdi}, {Botticella}, {Branchesi},
  {Brocato}, {Bruun}, {Bulla}, {Campana}, {Cappellaro}, {Castro-Tirado},
  {Chambers}, {Chaty}, {Chen}, {Ciolfi}, {Coleiro}, {Copperwheat}, {Covino},
  {Cutter}, {D'Ammand o}, {D'Avanzo}, {De Cesare}, {D'Elia}, {Della Valle},
  {Denneau}, {De Pasquale}, {Dhillon}, {Dyer}, {Elias-Rosa}, {Evans},
  {Eyles-Ferris}, {Fiore}, {Fraser}, {Fruchter}, {Fynbo}, {Galbany}, {Gall},
  {Galloway}, {Getman}, {Ghirlanda}, {Gillanders}, {Gomboc}, {Gompertz},
  {Gonz{\'a}lez-Fern{\'a}ndez}, {Gonz{\'a}lez-Gait{\'a}n}, {Grado}, {Greco},
  {Gromadzki}, {Groot}, {Guti{\'e}rrez}, {Heikkil{\"a}}, {Heintz}, {Hjorth},
  {Hu}, {Huber}, {Inserra}, {Izzo}, {Japelj}, {Jerkstrand}, {Jin}, {Jonker},
  {Kankare}, {Kann}, {Kennedy}, {Kim}, {Klose}, {Kool}, {Kotak},
  {Kuncarayakti}, {Lamb}, {Leloudas}, {Levan}, {Longo}, {Lowe}, {Lyman},
  {Magnier}, {Maguire}, {Maiorano}, {Mandel}, {Mapelli}, {Mattila}, {McBrien},
  {Melandri}, {Micha{\l}owski}, {Milvang-Jensen}, {Moran}, {Nicastro},
  {Nicholl}, {Nicuesa Guelbenzu}, {Nuttal}, {Oates}, {O'Brien}, {Onori},
  {Palazzi}, {Patricelli}, {Perego}, {Torres}, {Perley}, {Pian}, {Pignata},
  {Piranomonte}, {Poshyachinda}, {Possenti}, {Pumo}, {Quirola-V{\'a}squez},
  {Ragosta}, {Ramsay}, {Rau}, {Rest}, {Reynolds}, {Rosetti}, {Rossi},
  {Rosswog}, {Sabha}, {Sagu{\'e}s Carracedo}, {Salafia}, {Salmon},
  {Salvaterra}, {Savaglio}, {Sbordone}, {Schady}, {Schipani}, {Schultz},
  {Schweyer}, {Smartt}, {Smith}, {Smith}, {Sollerman}, {Srivastav}, {Stanway},
  {Starling}, {Steeghs}, {Stratta}, {Stubbs}, {Tanvir}, {Testa}, {Thrane},
  {Tonry}, {Turatto}, {Ulaczyk}, {van der Horst}, {Vergani}, {Walton},
  {Watson}, {Wiersema}, {Wiik}, {Wyrzykowski}, {Yang}, {Yi}, \&
  {Young}}]{Ackley2020}
{Ackley}, K., {Amati}, L., {Barbieri}, C., {et~al.} 2020, arXiv e-prints,
  arXiv:2002.01950

\bibitem[{{Adams} {et~al.}(2018){Adams}, {Boyajian}, \& {von
  Braun}}]{Adams2018}
{Adams}, A.~D., {Boyajian}, T.~S., \& {von Braun}, K. 2018, \mnras, 473, 3608

\bibitem[{{Albrow} {et~al.}(1999){Albrow}, {Beaulieu}, {Caldwell}, {DePoy},
  {Dominik}, {Gaudi}, {Gould}, {Greenhill}, {Hill}, {Kane}, {Martin},
  {Menzies}, {Naber}, {Pogge}, {Pollard}, {Sackett}, {Sahu}, {Vermaak},
  {Watson}, {Williams}, \& {PLANET Collaboration}}]{Albrow1999_LD}
{Albrow}, M.~D., {Beaulieu}, J.~P., {Caldwell}, J.~A.~R., {et~al.} 1999, \apj,
  522, 1022

\bibitem[{{Alcock} {et~al.}(2001){Alcock}, {Allsman}, {Alves}, {Axelrod},
  {Becker}, {Bennett}, {Cook}, {Drake}, {Freeman}, {Geha}, {Griest}, {Keller},
  {Lehner}, {Marshall}, {Minniti}, {Nelson}, {Peterson}, {Popowski}, {Pratt},
  {Quinn}, {Stubbs}, {Sutherland}, {Tomaney}, {Vandehei}, \&
  {Welch}}]{Alcock2001}
{Alcock}, C., {Allsman}, R.~A., {Alves}, D.~R., {et~al.} 2001, \nat, 414, 617

\bibitem[{{Bachelet} {et~al.}(2017){Bachelet}, {Norbury}, {Bozza}, \&
  {Street}}]{pyLIMA2017}
{Bachelet}, E., {Norbury}, M., {Bozza}, V., \& {Street}, R. 2017, \aj, 154, 203

\bibitem[{{Bachelet} {et~al.}(2021){Bachelet}, {Zieli{\'n}ski}, {Gromadzki},
  {Gezer}, {Rybicki}, \& {Kruszy{\'n}ska}}]{Bachelet2021}
{Bachelet}, E., {Zieli{\'n}ski}, P., {Gromadzki}, M., {et~al.} 2021, \aap

\bibitem[{{Bailer-Jones} {et~al.}(2020){Bailer-Jones}, {Rybizki}, {Fouesneau},
  {Demleitner}, \& {Andrae}}]{CBJ2020}
{Bailer-Jones}, C.~A.~L., {Rybizki}, J., {Fouesneau}, M., {Demleitner}, M., \&
  {Andrae}, R. 2020, arXiv e-prints, arXiv:2012.05220

\bibitem[{{Bennett} {et~al.}(2015){Bennett}, {Bhattacharya}, {Anderson},
  {Bond}, {Anderson}, {Barry}, {Batista}, {Beaulieu}, {DePoy}, {Dong}, {Gaudi},
  {Gilbert}, {Gould}, {Pfeifle}, {Pogge}, {Suzuki}, {Terry}, \&
  {Udalski}}]{Bennett2015}
{Bennett}, D.~P., {Bhattacharya}, A., {Anderson}, J., {et~al.} 2015, \apj, 808,
  169

\bibitem[{{Bennett} {et~al.}(2020){Bennett}, {Bhattacharya}, {Beaulieu},
  {Blackman}, {Vandorou}, {Terry}, {Cole}, {Henderson}, {Koshimoto}, {Lu},
  {Baptiste Marquette}, {Ranc}, \& {Udalski}}]{Bennett2020}
{Bennett}, D.~P., {Bhattacharya}, A., {Beaulieu}, J.-P., {et~al.} 2020, \aj,
  159, 68

\bibitem[{{Bertin}(2006)}]{scamp2006}
{Bertin}, E. 2006, in Astronomical Society of the Pacific Conference Series,
  Vol. 351, Astronomical Data Analysis Software and Systems XV, ed.
  C.~{Gabriel}, C.~{Arviset}, D.~{Ponz}, \& S.~{Enrique}, 112

\bibitem[{{Bertin} \& {Arnouts}(1996)}]{sextractor1996}
{Bertin}, E. \& {Arnouts}, S. 1996, \aaps, 117, 393

\bibitem[{{Bhattacharya} {et~al.}(2018){Bhattacharya}, {Beaulieu}, {Bennett},
  {Anderson}, {Koshimoto}, {Lu}, {Batista}, {Blackman}, {Bond}, {Fukui},
  {Henderson}, {Hirao}, {Marquette}, {Mroz}, {Ranc}, \&
  {Udalski}}]{Bhattacharya2018}
{Bhattacharya}, A., {Beaulieu}, J.~P., {Bennett}, D.~P., {et~al.} 2018, \aj,
  156, 289

\bibitem[{{Bond} {et~al.}(2001){Bond}, {Abe}, {Dodd}, {Hearnshaw}, {Honda},
  {Jugaku}, {Kilmartin}, {Marles}, {Masuda}, {Matsubara}, {Muraki}, {Nakamura},
  {Nankivell}, {Noda}, {Noguchi}, {Ohnishi}, {Rattenbury}, {Reid}, {Saito},
  {Sato}, {Sekiguchi}, {Skuljan}, {Sullivan}, {Sumi}, {Takeuti}, {Watase},
  {Wilkinson}, {Yamada}, {Yanagisawa}, \& {Yock}}]{MOA2001}
{Bond}, I.~A., {Abe}, F., {Dodd}, R.~J., {et~al.} 2001, \mnras, 327, 868

\bibitem[{{Boyajian} {et~al.}(2014){Boyajian}, {van Belle}, \& {von
  Braun}}]{Boyajian2014}
{Boyajian}, T.~S., {van Belle}, G., \& {von Braun}, K. 2014, \aj, 147, 47

\bibitem[{{Calchi Novati} {et~al.}(2015{\natexlab{a}}){Calchi Novati}, {Gould},
  {Udalski}, {Menzies}, {Bond}, {Shvartzvald}, {Street}, {Hundertmark},
  {Beichman}, {Yee}, {Carey}, {Poleski}, {Skowron}, {Koz{\l}owski}, {Mr{\'o}z},
  {Pietrukowicz}, {Pietrzy{\'n}ski}, {Szyma{\'n}ski}, {Soszy{\'n}ski},
  {Ulaczyk}, {Wyrzykowski}, {OGLE Collaboration}, {Albrow}, {Beaulieu},
  {Caldwell}, {Cassan}, {Coutures}, {Danielski}, {Dominis Prester},
  {Donatowicz}, {Lon{\v c}ari{\'c}}, {McDougall}, {Morales}, {Ranc}, {Zhu},
  {PLANET Collaboration}, {Abe}, {Barry}, {Bennett}, {Bhattacharya},
  {Fukunaga}, {Inayama}, {Koshimoto}, {Namba}, {Sumi}, {Suzuki}, {Tristram},
  {Wakiyama}, {Yonehara}, {MOA Collaboration}, {Maoz}, {Kaspi}, {Friedmann},
  {Wise Group}, {Bachelet}, {Figuera Jaimes}, {Bramich}, {Tsapras}, {Horne},
  {Snodgrass}, {Wambsganss}, {Steele}, {Kains}, {RoboNet Collaboration},
  {Bozza}, {Dominik}, {J{\o}rgensen}, {Alsubai}, {Ciceri}, {D'Ago},
  {Haugb{\o}lle}, {Hessman}, {Hinse}, {Juncher}, {Korhonen}, {Mancini},
  {Popovas}, {Rabus}, {Rahvar}, {Scarpetta}, {Schmidt}, {Skottfelt},
  {Southworth}, {Starkey}, {Surdej}, {Wertz}, {Zarucki}, {MiNDSTEp Consortium},
  {Gaudi}, {Pogge}, {DePoy}, \& {{$\mu$}FUN Collaboration}}]{CalchiNovati2015}
{Calchi Novati}, S., {Gould}, A., {Udalski}, A., {et~al.} 2015{\natexlab{a}},
  \apj, 804, 20

\bibitem[{{Calchi Novati} {et~al.}(2015{\natexlab{b}}){Calchi Novati}, {Gould},
  {Yee}, {Beichman}, {Bryden}, {Carey}, {Fausnaugh}, {Gaudi}, {Henderson},
  {Pogge}, {Shvartzvald}, {Wibking}, {Zhu}, {Spitzer Team}, {Udalski},
  {Poleski}, {Pawlak}, {Szyma{\'n}ski}, {Skowron}, {Mr{\'o}z}, {Koz{\l}owski},
  {Wyrzykowski}, {Pietrukowicz}, {Pietrzy{\'n}ski}, {Soszy{\'n}ski}, {Ulaczyk},
  \& {OGLE Group}}]{SpitzerReduction2015}
{Calchi Novati}, S., {Gould}, A., {Yee}, J.~C., {et~al.} 2015{\natexlab{b}},
  \apj, 814, 92

\bibitem[{{Campbell} {et~al.}(2015){Campbell}, {Marsh}, {Fraser}, {Hodgkin},
  {de Miguel}, {G{\"a}nsicke}, {Steeghs}, {Hourihane}, {Breedt}, {Littlefair},
  {Koposov}, {Wyrzykowski}, {Altavilla}, {Blagorodnova}, {Clementini},
  {Damljanovic}, {Delgado}, {Dennefeld}, {Drake},
  {Fern{\'a}ndez-Hern{\'a}ndez}, {Gilmore}, {Gualandi}, {Hamanowicz},
  {Handzlik}, {Hardy}, {Harrison}, {I{\l}kiewicz}, {Jonker}, {Kochanek},
  {Ko{\l}aczkowski}, {Kostrzewa-Rutkowska}, {Kotak}, {van Leeuwen}, {Leto},
  {Ochner}, {Pawlak}, {Palaversa}, {Rixon}, {Rybicki}, {Shappee}, {Smartt},
  {Torres}, {Tomasella}, {Turatto}, {Ulaczyk}, {van Velzen}, {Vince}, {Walton},
  {Wielg{\'o}rski}, {Wevers}, {Whitelock}, {Yoldas}, {De Angeli}, {Burgess},
  {Busso}, {Busuttil}, {Butterley}, {Chambers}, {Copperwheat}, {Danilet},
  {Dhillon}, {Evans}, {Eyer}, {Froebrich}, {Gomboc}, {Holland}, {Holoien},
  {Jarvis}, {Kaiser}, {Kann}, {Koester}, {Kolb}, {Komossa}, {Magnier},
  {Mahabal}, {Polshaw}, {Prieto}, {Prusti}, {Riello}, {Scholz}, {Simonian},
  {Stanek}, {Szabados}, {Waters}, \& {Wilson}}]{Campbell2015}
{Campbell}, H.~C., {Marsh}, T.~R., {Fraser}, M., {et~al.} 2015, \mnras, 452,
  1060

\bibitem[{{Carr} {et~al.}(2021){Carr}, {Clesse}, \&
  {Garc{\'\i}a-Bellido}}]{Carr2021}
{Carr}, B., {Clesse}, S., \& {Garc{\'\i}a-Bellido}, J. 2021, \mnras, 501, 1426

\bibitem[{{Carr} \& {Silk}(2018)}]{Carr2018}
{Carr}, B. \& {Silk}, J. 2018, \mnras, 478, 3756

\bibitem[{{Cassan} \& {Ranc}(2016)}]{Cassan2016}
{Cassan}, A. \& {Ranc}, C. 2016, \mnras, 458, 2074

\bibitem[{{Cassan} {et~al.}(2021){Cassan}, {Ranc}, {Absil}, {Wyrzykowski},
  {Rybicki}, \& {Bachelet}}]{Cassan2021}
{Cassan}, A., {Ranc}, C., {Absil}, O., {et~al.} 2021, Nature Astronomy

\bibitem[{{Claret}(2000)}]{Claret2000}
{Claret}, A. 2000, \aap, 363, 1081

\bibitem[{{Claret}(2019)}]{Claret2019}
{Claret}, A. 2019, Research Notes of the American Astronomical Society, 3, 17

\bibitem[{{Delgado} {et~al.}(2019){Delgado}, {Harrison}, {Hodgkin}, {Leeuwen},
  {Rixon}, \& {Yoldas}}]{Gaia19bldTNS}
{Delgado}, A., {Harrison}, D., {Hodgkin}, S., {et~al.} 2019, Transient Name
  Server Discovery Report, 2019-605, 1

\bibitem[{{Delplancke} {et~al.}(2001){Delplancke}, {G{\'o}rski}, \&
  {Richichi}}]{Delplancke2001}
{Delplancke}, F., {G{\'o}rski}, K.~M., \& {Richichi}, A. 2001, \aap, 375, 701

\bibitem[{{Dominik} \& {Sahu}(2000)}]{Dominik2000}
{Dominik}, M. \& {Sahu}, K.~C. 2000, \apj, 534, 213

\bibitem[{{Dong} {et~al.}(2019){Dong}, {M{\'e}rand}, {Delplancke-Str{\"o}bele},
  {Gould}, {Chen}, {Post}, {Kochanek}, {Stanek}, {Christie}, {Mutel},
  {Natusch}, {Holoien}, {Prieto}, {Shappee}, \& {Thompson}}]{Dong2019}
{Dong}, S., {M{\'e}rand}, A., {Delplancke-Str{\"o}bele}, F., {et~al.} 2019,
  \apj, 871, 70

\bibitem[{{Einstein}(1936)}]{1936Einstein}
{Einstein}, A. 1936, Science, 84, 506

\bibitem[{{Evans} {et~al.}(2018){Evans}, {Riello}, {De Angeli}, {Carrasco},
  {Montegriffo}, {Fabricius}, {Jordi}, {Palaversa}, {Diener}, {Busso},
  {Cacciari}, {van Leeuwen}, {Burgess}, {Davidson}, {Harrison}, {Hodgkin},
  {Pancino}, {Richards}, {Altavilla}, {Balaguer-N{\'u}{\~n}ez}, {Barstow},
  {Bellazzini}, {Brown}, {Castellani}, {Cocozza}, {De Luise}, {Delgado},
  {Ducourant}, {Galleti}, {Gilmore}, {Giuffrida}, {Holl}, {Kewley}, {Koposov},
  {Marinoni}, {Marrese}, {Osborne}, {Piersimoni}, {Portell}, {Pulone},
  {Ragaini}, {Sanna}, {Terrett}, {Walton}, {Wevers}, \&
  {Wyrzykowski}}]{GaiaDR2photo}
{Evans}, D.~W., {Riello}, M., {De Angeli}, F., {et~al.} 2018, \aap, 616, A4

\bibitem[{{Foreman-Mackey} {et~al.}(2013){Foreman-Mackey}, {Hogg}, {Lang}, \&
  {Goodman}}]{EMCEE}
{Foreman-Mackey}, D., {Hogg}, D.~W., {Lang}, D., \& {Goodman}, J. 2013, \pasp,
  125, 306

\bibitem[{{Fukui} {et~al.}(2019){Fukui}, {Suzuki}, {Koshimoto}, {Bachelet},
  {Vanmunster}, {Storey}, {Maehara}, {Yanagisawa}, {Yamada}, {Yonehara},
  {Hirano}, {Bennett}, {Bozza}, {Mawet}, {Penny}, {Awiphan}, {Oksanen},
  {Heintz}, {Oberst}, {B{\'e}jar}, {Casasayas-Barris}, {Chen}, {Crouzet},
  {Hidalgo}, {Klagyivik}, {Murgas}, {Narita}, {Palle}, {Parviainen},
  {Watanabe}, {Kusakabe}, {Mori}, {Terada}, {de Leon}, {Hernandez}, {Luque},
  {Monelli}, {Monta{\~n}es-Rodriguez}, {Prieto-Arranz}, {Murata}, {Shugarov},
  {Kubota}, {Otsuki}, {Shionoya}, {Nishiumi}, {Nishide}, {Fukagawa}, {Onodera},
  {Villanueva}, {Street}, {Tsapras}, {Hundertmark}, {Kuzuhara}, {Fujita},
  {Beichman}, {Beaulieu}, {Alonso}, {Reichart}, {Kawai}, \&
  {Tamura}}]{Fukui2019}
{Fukui}, A., {Suzuki}, D., {Koshimoto}, N., {et~al.} 2019, \aj, 158, 206

\bibitem[{{Garc{\'{\i}}a-Bellido} {et~al.}(2018){Garc{\'{\i}}a-Bellido},
  {Clesse}, \& {Fleury}}]{Garcia-Bellido2018}
{Garc{\'{\i}}a-Bellido}, J., {Clesse}, S., \& {Fleury}, P. 2018, Physics of the
  Dark Universe, 20, 95

\bibitem[{{Gillessen} {et~al.}(2012){Gillessen}, {Genzel}, {Fritz}, {Quataert},
  {Alig}, {Burkert}, {Cuadra}, {Eisenhauer}, {Pfuhl}, {Dodds-Eden}, {Gammie},
  \& {Ott}}]{Gillessen2012}
{Gillessen}, S., {Genzel}, R., {Fritz}, T.~K., {et~al.} 2012, \nat, 481, 51

\bibitem[{{Gilliland} {et~al.}(2010){Gilliland}, {Brown},
  {Christensen-Dalsgaard}, {Kjeldsen}, {Aerts}, {Appourchaux}, {Basu},
  {Bedding}, {Chaplin}, {Cunha}, {De Cat}, {De Ridder}, {Guzik}, {Handler},
  {Kawaler}, {Kiss}, {Kolenberg}, {Kurtz}, {Metcalfe}, {Monteiro}, {Szab{\'o}},
  {Arentoft}, {Balona}, {Debosscher}, {Elsworth}, {Quirion}, {Stello},
  {Su{\'a}rez}, {Borucki}, {Jenkins}, {Koch}, {Kondo}, {Latham}, {Rowe}, \&
  {Steffen}}]{Gilliland2010}
{Gilliland}, R.~L., {Brown}, T.~M., {Christensen-Dalsgaard}, J., {et~al.} 2010,
  \pasp, 122, 131

\bibitem[{{Gould}(1992)}]{Gould1992}
{Gould}, A. 1992, \apj, 392, 442

\bibitem[{{Gould}(1994{\natexlab{a}})}]{Gould1994_space}
{Gould}, A. 1994{\natexlab{a}}, \apjl, 421, L75

\bibitem[{{Gould}(1994{\natexlab{b}})}]{Gould1994}
{Gould}, A. 1994{\natexlab{b}}, \apjl, 421, L71

\bibitem[{{Gould}(2000)}]{Gould2000b}
{Gould}, A. 2000, \apj, 542, 785

\bibitem[{{Gould}(2004)}]{Gould2004}
{Gould}, A. 2004, \apj, 606, 319

\bibitem[{{Gould} {et~al.}(2020){Gould}, {Ryu}, {Calchi Novati}, {Zang},
  {Albrow}, {Chung}, {Han}, {Hwang}, {Jung}, {Shin}, {Shvartzvald}, {Yee},
  {Cha}, {Kim}, {Kim}, {Kim}, {Lee}, {Lee}, {Lee}, {Park}, {Pogge}, {Beichman},
  {Bryden}, {Carey}, {Gaudi}, {Henderson}, {Zhu}, {Fouque}, {Penny}, {Petric},
  {Burdullis}, \& {Mao}}]{Gould2020}
{Gould}, A., {Ryu}, Y.-H., {Calchi Novati}, S., {et~al.} 2020, Journal of
  Korean Astronomical Society, 53, 9

\bibitem[{{Gould} \& {Yee}(2014)}]{GouldYee2014}
{Gould}, A. \& {Yee}, J.~C. 2014, \apj, 784, 64

\bibitem[{{Graczyk} {et~al.}(2018){Graczyk}, {Pietrzy{\'n}ski}, {Thompson},
  {Gieren}, {Pilecki}, {Konorski}, {Villanova}, {G{\'o}rski}, {Suchomska},
  {Karczmarek}, {Stepie{\'n}}, {Storm}, {Taormina}, {Ko{\l}aczkowski},
  {Wielg{\'o}rski}, {Narloch}, {Zgirski}, {Gallenne}, {Ostrowski}, {Smolec},
  {Udalski}, {Soszy{\'n}ski}, {Kervella}, {Nardetto}, {Szyma{\'n}ski},
  {Wyrzykowski}, {Ulaczyk}, {Poleski}, {Pietrukowicz}, {Koz{\l}owski},
  {Skowron}, \& {Mr{\'o}z}}]{Graczyk2018}
{Graczyk}, D., {Pietrzy{\'n}ski}, G., {Thompson}, I.~B., {et~al.} 2018, \apj,
  860, 1

\bibitem[{{Hirao} {et~al.}(2020){Hirao}, {Bennett}, {Ryu}, {Koshimoto},
  {Udalski}, {Yee}, {Sumi}, {Bond}, {Shvartzvald}, {Abe}, {Barry},
  {Bhattacharya}, {Donachie}, {Fukui}, {Itow}, {Kondo}, {Man Cheung},
  {Matsubara}, {Matsuo}, {Miyazaki}, {Muraki}, {Nagakane}, {Ranc},
  {Rattenbury}, {Suematsu}, {Shibai}, {Suzuki}, {Tristram}, {Yonehara}, {MOA
  Collaboration}, {Skowron}, {Poleski}, {Mr{\'o}z}, {Szyma{\'n}ski},
  {Soszy{\'n}ski}, {Koz{\l}owski}, {Pietrukowicz}, {Ulaczyk}, {Rybicki},
  {Iwanek}, {OGLE Collaboration}, {Albrow}, {Chung}, {Gould}, {Han}, {Hwang},
  {Jung}, {Shin}, {Zang}, {Cha}, {Kim}, {Kim}, {Kim}, {Lee}, {Lee}, {Lee},
  {Park}, {Pogge}, {KMTNet Collaboration}, {Beichman}, {Bryden}, {Novati},
  {Carey}, {Gaudi}, {Henderson}, {Zhu}, {Spitzer Team}, {Bachelet}, {Bolt},
  {Christie}, {Hundertmark}, {Natusch}, {Maoz}, {McCormick}, {Street}, {Tan},
  {Tsapras}, {LCO}, {{\ensuremath{\mu}}FUN Follow-up Teams}, {J{\o}rgensen},
  {Dominik}, {Bozza}, {Skottfelt}, {Snodgrass}, {Ciceri}, {Jaimes}, {Evans},
  {Peixinho}, {Hinse}, {Burgdorf}, {Southworth}, {Rahvar}, {Sajadian}, {Rabus},
  {von Essen}, {Fujii}, {Campbell-White}, {Lowry}, {Helling}, {Mancini},
  {Haikala}, {MindSTEp Collaboration}, {Kandori}, \& {IRSF Team}}]{Hirao2020}
{Hirao}, Y., {Bennett}, D.~P., {Ryu}, Y.-H., {et~al.} 2020, \aj, 160, 74

\bibitem[{{Hodgkin} {et~al.}(2021){Hodgkin}, {Harrison}, {Breedt}, {Wevers},
  {Rixon}, {Delgado}, {Yoldas}, {Kostrzewa-Rutkowska}, {Wyrzykowski}, {van
  Leeuwen}, {Blagorodnova}, {Campbell}, {Eappachen}, {Fraser}, {Ihanec},
  {Koposov}, {Kruszy{\'n}ska}, {Marton}, {Rybicki}, {Brown}, {Burgess},
  {Busso}, {Cowell}, {De Angeli}, {Diener}, {Evans}, {Gilmore}, {Holland},
  {Jonker}, {van Leeuwen}, {Mignard}, {Osborne}, {Portell}, {Prusti},
  {Richards}, {Riello}, {Seabroke}, {Walton}, {{\'A}brah{\'a}m}, {Altavilla},
  {Baker}, {Bastian}, {O'Brien}, {de Bruijne}, {Butterley}, {Carrasco},
  {Casta{\~n}eda}, {Clark}, {Clementini}, {Copperwheat}, {Cropper},
  {Damljanovic}, {Davidson}, {Davis}, {Dennefeld}, {Dhillon}, {Dolding},
  {Dominik}, {Esquej}, {Eyer}, {Fabricius}, {Fridman}, {Froebrich}, {Garralda},
  {Gomboc}, {Gonz{\'a}lez-Vidal}, {Guerra}, {Hambly}, {Hardy}, {Holl},
  {Hourihane}, {Japelj}, {Kann}, {Kiss}, {Knigge}, {Kolb}, {Komossa},
  {K{\'o}sp{\'a}l}, {Kov{\'a}cs}, {Kun}, {Leto}, {Lewis}, {Littlefair},
  {Mahabal}, {Mundell}, {Nagy}, {Padeletti}, {Palaversa}, {Pigulski},
  {Pretorius}, {van Reeven}, {Ribeiro}, {Roelens}, {Rowell}, {Schartel},
  {Scholz}, {Schwope}, {Sip{\H{o}}cz}, {Smartt}, {Smith}, {Serraller},
  {Steeghs}, {Sullivan}, {Szabados}, {Szegedi-Elek}, {Tisserand}, {Tomasella},
  {van Velzen}, {Whitelock}, {Wilson}, \& {Young}}]{Hodgkin2021}
{Hodgkin}, S.~T., {Harrison}, D.~L., {Breedt}, E., {et~al.} 2021, \aap, 652,
  A76

\bibitem[{{Hodgkin} {et~al.}(2013){Hodgkin}, {Wyrzykowski}, {Blagorodnova}, \&
  {Koposov}}]{Hodgkin2013}
{Hodgkin}, S.~T., {Wyrzykowski}, L., {Blagorodnova}, N., \& {Koposov}, S. 2013,
  Philosophical Transactions of the Royal Society of London Series A, 371,
  20120239

\bibitem[{{H{\o}g}(1995)}]{Hog1995}
{H{\o}g}, E. 1995, in ESA Special Publication, Vol. 379, Future Possibilities
  for bstrometry in Space, ed. M.~A.~C. {Perryman} \& F.~{van Leeuwen}, 125

\bibitem[{{Holtzman} {et~al.}(1998){Holtzman}, {Watson}, {Baum}, {Grillmair},
  {Groth}, {Light}, {Lynds}, \& {O'Neil}}]{holtzman98}
{Holtzman}, J.~A., {Watson}, A.~M., {Baum}, W.~A., {et~al.} 1998, \aj, 115,
  1946

\bibitem[{{Kallinger} {et~al.}(2010){Kallinger}, {Weiss}, {Barban}, {Baudin},
  {Cameron}, {Carrier}, {De Ridder}, {Goupil}, {Gruberbauer}, {Hatzes},
  {Hekker}, {Samadi}, \& {Deleuil}}]{Kallinger2010}
{Kallinger}, T., {Weiss}, W.~W., {Barban}, C., {et~al.} 2010, \aap, 509, A77

\bibitem[{{Kim} {et~al.}(2016){Kim}, {Lee}, {Park}, {Kim}, {Cha}, {Lee}, {Han},
  {Chun}, \& {Yuk}}]{Kim2016}
{Kim}, S.-L., {Lee}, C.-U., {Park}, B.-G., {et~al.} 2016, Journal of Korean
  Astronomical Society, 49, 37

\bibitem[{{Kiziltan} {et~al.}(2013){Kiziltan}, {Kottas}, {De Yoreo}, \&
  {Thorsett}}]{Kiziltan2013}
{Kiziltan}, B., {Kottas}, A., {De Yoreo}, M., \& {Thorsett}, S.~E. 2013, \apj,
  778, 66

\bibitem[{{Kjeldsen} \& {Bedding}(1995)}]{Kjeldsen1995}
{Kjeldsen}, H. \& {Bedding}, T.~R. 1995, \aap, 293, 87

\bibitem[{{Kl{\"u}ter} {et~al.}(2019){Kl{\"u}ter}, {Bastian}, \&
  {Wambsganss}}]{Kluter2019}
{Kl{\"u}ter}, J., {Bastian}, U., \& {Wambsganss}, J. 2019, arXiv e-prints,
  arXiv:1911.02584

\bibitem[{{Le Bouquin} {et~al.}(2017){Le Bouquin}, {Sana}, {Gosset}, {De
  Becker}, {Duvert}, {Absil}, {Anthonioz}, {Berger}, {Ertel}, {Grellmann},
  {Guieu}, {Kervella}, {Rabus}, \& {Willson}}]{LeBouquin2017}
{Le Bouquin}, J.~B., {Sana}, H., {Gosset}, E., {et~al.} 2017, \aap, 601, A34

\bibitem[{{Lee} {et~al.}(2010){Lee}, {Seitz}, {Riffeser}, \&
  {Bender}}]{Lee_FS_astro2010}
{Lee}, C.~H., {Seitz}, S., {Riffeser}, A., \& {Bender}, R. 2010, \mnras, 407,
  1597

\bibitem[{{Li} {et~al.}(2019){Li}, {Zang}, {Udalski}, {Shvartzvald}, {Huber},
  {Lee}, {Sumi}, {Gould}, {Mao}, {Fouqu{\'e}}, {Wang}, {Dong}, {J{\o}rgensen},
  {Cole}, {Mr{\'o}z}, {Szyma{\'n}ski}, {Skowron}, {Poleski}, {Soszy{\'n}ski},
  {Pietrukowicz}, {Koz{\l}owski}, {Ulaczyk}, {Rybicki}, {Iwanek}, {Yee},
  {Calchi Novati}, {Beichman}, {Bryden}, {Carey}, {Gaudi}, {Henderson}, {Zhu},
  {Albrow}, {Chung}, {Han}, {Hwang}, {Jung}, {Ryu}, {Shin}, {Cha}, {Kim},
  {Kim}, {Kim}, {Lee}, {Lee}, {Park}, {Pogge}, {Bond}, {Abe}, {Barry},
  {Bennett}, {Bhattacharya}, {Donachie}, {Fukui}, {Hirao}, {Itow}, {Kondo},
  {Koshimoto}, {Li}, {Matsubara}, {Muraki}, {Miyazaki}, {Nagakane}, {Ranc},
  {Rattenbury}, {Suematsu}, {Sullivan}, {Suzuki}, {Tristram}, {Yonehara},
  {Christie}, {Drummond}, {Green}, {Hennerley}, {Natusch}, {Porritt},
  {Bachelet}, {Maoz}, {Street}, {Tsapras}, {Bozza}, {Dominik}, {Hundertmark},
  {Peixinho}, {Sajadian}, {Burgdorf}, {Evans}, {Figuera Jaimes}, {Fujii},
  {Haikala}, {Helling}, {Henning}, {Hinse}, {Mancini}, {Longa-Pe{\~n}a},
  {Rahvar}, {Rabus}, {Skottfelt}, {Snodgrass}, {Southworth}, {Unda-Sanzana},
  {von Essen}, {Beaulieu}, {Blackman}, \& {Hill}}]{Li2019}
{Li}, S.~S., {Zang}, W., {Udalski}, A., {et~al.} 2019, \mnras, 488, 3308

\bibitem[{{Liebes}(1964)}]{Liebes1964}
{Liebes}, S. 1964, Physical Review, 133, 835

\bibitem[{{Miyamoto} \& {Yoshii}(1995)}]{Miyamoto1995}
{Miyamoto}, M. \& {Yoshii}, Y. 1995, \aj, 110, 1427

\bibitem[{{Mr{\'o}z} {et~al.}(2020){Mr{\'o}z}, {Udalski}, {Szyma{\'n}ski},
  {Soszy{\'n}ski}, {Pietrukowicz}, {Koz{\l}owski}, {Skowron}, {Poleski},
  {Ulaczyk}, {Gromadzki}, {Rybicki}, {Iwanek}, \& {Wrona}}]{Mroz2020}
{Mr{\'o}z}, P., {Udalski}, A., {Szyma{\'n}ski}, M.~K., {et~al.} 2020, \apjs,
  249, 16

\bibitem[{{Nemiroff} \& {Wickramasinghe}(1994)}]{Nemiroff1994}
{Nemiroff}, R.~J. \& {Wickramasinghe}, W.~A.~D.~T. 1994, \apjl, 424, L21

\bibitem[{{Nucita} {et~al.}(2018){Nucita}, {Licchelli}, {De Paolis},
  {Ingrosso}, {Strafella}, {Katysheva}, \& {Shugarov}}]{Nucita2018}
{Nucita}, A.~A., {Licchelli}, D., {De Paolis}, F., {et~al.} 2018, \mnras, 476,
  2962

\bibitem[{{Olejak} {et~al.}(2020){Olejak}, {Belczynski}, {Bulik}, \&
  {Sobolewska}}]{Olejak2020}
{Olejak}, A., {Belczynski}, K., {Bulik}, T., \& {Sobolewska}, M. 2020, \aap,
  638, A94

\bibitem[{{{\"O}zel} {et~al.}(2010){{\"O}zel}, {Psaltis}, {Narayan}, \&
  {McClintock}}]{Ozel2010}
{{\"O}zel}, F., {Psaltis}, D., {Narayan}, R., \& {McClintock}, J.~E. 2010,
  \apj, 725, 1918

\bibitem[{{Paczynski}(1986)}]{Paczynski1986}
{Paczynski}, B. 1986, \apj, 304, 1

\bibitem[{{Park} {et~al.}(2004){Park}, {DePoy}, {Gaudi}, {Gould}, {Han},
  {Pogge}, {muFun Collaboration}, {Abe}, {Bennett}, {Bond}, {Eguchi}, {Furuta},
  {Hearnshaw}, {Kamiya}, {Kilmartin}, {Kurata}, {Masuda}, {Matsubara},
  {Muraki}, {Noda}, {Okajima}, {Rattenbury}, {Sako}, {Sekiguchi}, {Sullivan},
  {Sumi}, {Tristram}, {Yanagisawa}, {Yock}, \& {MOA Collaboration}}]{mb03037}
{Park}, B.~G., {DePoy}, D.~L., {Gaudi}, B.~S., {et~al.} 2004, \apj, 609, 166

\bibitem[{{Pecaut} {et~al.}(2012){Pecaut}, {Mamajek}, \& {Bubar}}]{Mamajek1}
{Pecaut}, M.~J., {Mamajek}, E.~E., \& {Bubar}, E.~J. 2012, \apj, 746, 154

\bibitem[{{Penny} {et~al.}(2019){Penny}, {Gaudi}, {Kerins}, {Rattenbury},
  {Mao}, {Robin}, \& {Calchi Novati}}]{Penny2019}
{Penny}, M.~T., {Gaudi}, B.~S., {Kerins}, E., {et~al.} 2019, \apjs, 241, 3

\bibitem[{{Refsdal}(1964)}]{Refsdal964}
{Refsdal}, S. 1964, \mnras, 128, 295

\bibitem[{{Refsdal}(1966)}]{Refsdal1966}
{Refsdal}, S. 1966, \mnras, 134, 315

\bibitem[{{Robin} {et~al.}(2003){Robin}, {Reyl{\'e}}, {Derri{\`e}re}, \&
  {Picaud}}]{Besancon}
{Robin}, A.~C., {Reyl{\'e}}, C., {Derri{\`e}re}, S., \& {Picaud}, S. 2003,
  \aap, 409, 523

\bibitem[{{Rybicki} {et~al.}(2019){Rybicki}, {Wyrzykowski}, {Zielinski},
  {Ratajczak}, {Bachelet}, {Street}, {Hundertmark}, {Tsapras}, {Udalski},
  {Hambsch}, {Konacki}, {Pawlaszek}, \& {Kinasz}}]{Gaia19bldATel}
{Rybicki}, K., {Wyrzykowski}, L., {Zielinski}, P., {et~al.} 2019, The
  Astronomer's Telegram, 12948, 1

\bibitem[{{Rybicki} {et~al.}(2018){Rybicki}, {Wyrzykowski}, {Klencki}, {de
  Bruijne}, {Belczy{\'n}ski}, \& {Chru{\'s}li{\'n}ska}}]{Rybicki2018}
{Rybicki}, K.~A., {Wyrzykowski}, {\L}., {Klencki}, J., {et~al.} 2018, \mnras,
  476, 2013

\bibitem[{{Sahu} {et~al.}(2017){Sahu}, {Anderson}, {Casertano}, {Bond},
  {Bergeron}, {Nelan}, {Pueyo}, {Brown}, {Bellini}, {Levay}, {Sokol}, {aff1},
  {Dominik}, {Calamida}, {Kains}, \& {Livio}}]{Sahu2017}
{Sahu}, K.~C., {Anderson}, J., {Casertano}, S., {et~al.} 2017, Science, 356,
  1046

\bibitem[{{Smartt} {et~al.}(2017){Smartt}, {Chen}, {Jerkstrand}, {Coughlin},
  {Kankare}, {Sim}, {Fraser}, {Inserra}, {Maguire}, {Chambers}, {Huber},
  {Kr{\"u}hler}, {Leloudas}, {Magee}, {Shingles}, {Smith}, {Young}, {Tonry},
  {Kotak}, {Gal-Yam}, {Lyman}, {Homan}, {Agliozzo}, {Anderson}, {Angus},
  {Ashall}, {Barbarino}, {Bauer}, {Berton}, {Botticella}, {Bulla}, {Bulger},
  {Cannizzaro}, {Cano}, {Cartier}, {Cikota}, {Clark}, {De Cia}, {Della Valle},
  {Denneau}, {Dennefeld}, {Dessart}, {Dimitriadis}, {Elias-Rosa}, {Firth},
  {Flewelling}, {Fl{\"o}rs}, {Franckowiak}, {Frohmaier}, {Galbany},
  {Gonz{\'a}lez-Gait{\'a}n}, {Greiner}, {Gromadzki}, {Guelbenzu},
  {Guti{\'e}rrez}, {Hamanowicz}, {Hanlon}, {Harmanen}, {Heintz}, {Heinze},
  {Hernandez}, {Hodgkin}, {Hook}, {Izzo}, {James}, {Jonker}, {Kerzendorf},
  {Klose}, {Kostrzewa-Rutkowska}, {Kowalski}, {Kromer}, {Kuncarayakti},
  {Lawrence}, {Lowe}, {Magnier}, {Manulis}, {Martin-Carrillo}, {Mattila},
  {McBrien}, {M{\"u}ller}, {Nordin}, {O'Neill}, {Onori}, {Palmerio},
  {Pastorello}, {Patat}, {Pignata}, {Podsiadlowski}, {Pumo}, {Prentice}, {Rau},
  {Razza}, {Rest}, {Reynolds}, {Roy}, {Ruiter}, {Rybicki}, {Salmon}, {Schady},
  {Schultz}, {Schweyer}, {Seitenzahl}, {Smith}, {Sollerman}, {Stalder},
  {Stubbs}, {Sullivan}, {Szegedi}, {Taddia}, {Taubenberger}, {Terreran}, {van
  Soelen}, {Vos}, {Wainscoat}, {Walton}, {Waters}, {Weiland}, {Willman},
  {Wiseman}, {Wright}, {Wyrzykowski}, \& {Yaron}}]{Smartt2017Nature}
{Smartt}, S.~J., {Chen}, T.-W., {Jerkstrand}, A., {et~al.} 2017, \nat, 551, 75

\bibitem[{{Smith} {et~al.}(2003){Smith}, {Mao}, \& {Paczy{\'n}ski}}]{Smith2003}
{Smith}, M.~C., {Mao}, S., \& {Paczy{\'n}ski}, B. 2003, \mnras, 339, 925

\bibitem[{{Stetson}(1987)}]{Stetson1987}
{Stetson}, P.~B. 1987, \pasp, 99, 191

\bibitem[{{Szegedi-Elek} {et~al.}(2020){Szegedi-Elek}, {{\'A}brah{\'a}m},
  {Wyrzykowski}, {Kun}, {K{\'o}sp{\'a}l}, {Chen}, {Marton}, {Mo{\'o}r}, {Kiss},
  {P{\'a}l}, {Szabados}, {Varga}, {Varga-Vereb{\'e}lyi}, {Andreas}, {Bachelet},
  {Bischoff}, {B{\'o}di}, {Breedt}, {Burgaz}, {Butterley}, {{\v{C}}epas},
  {Damljanovic}, {Gezer}, {Godunova}, {Gromadzki}, {Gurgul}, {Hardy},
  {Hildebrandt}, {Hoffmann}, {Hundertmark}, {Ihanec}, {Janulis}, {Kalup},
  {Kaczmarek}, {K{\"o}nyves-T{\'o}th}, {Krezinger}, {Kruszy{\'n}ska},
  {Littlefair}, {Maskoli{\={u}}nas}, {M{\'e}sz{\'a}ros}, {Miko{\l}ajczyk},
  {Mugrauer}, {Netzel}, {Ordasi}, {Pak{\v{s}}tien{\.{e}}}, {Rybicki},
  {S{\'a}rneczky}, {Seli}, {Simon}, {{\v{S}}i{\v{s}}kauskait{\.{e}}},
  {S{\'o}dor}, {Sokolovsky}, {Stenglein}, {Street}, {Szak{\'a}ts}, {Tomasella},
  {Tsapras}, {Vida}, {Zdana{\v{c}}ius}, {Zieli{\'n}ski}, {Zieli{\'n}ski}, \&
  {Zi{\'o}{\l}kowska}}]{Szegedi-Elek2020}
{Szegedi-Elek}, E., {{\'A}brah{\'a}m}, P., {Wyrzykowski}, L., {et~al.} 2020,
  arXiv e-prints, arXiv:2005.11537

\bibitem[{{Udalski} {et~al.}(1992){Udalski}, {Szymanski}, {Kaluzny}, {Kubiak},
  \& {Mateo}}]{Udalski1992}
{Udalski}, A., {Szymanski}, M., {Kaluzny}, J., {Kubiak}, M., \& {Mateo}, M.
  1992, \actaa, 42, 253

\bibitem[{{Udalski} {et~al.}(2015{\natexlab{a}}){Udalski}, {Szyma{\'n}ski}, \&
  {Szyma{\'n}ski}}]{Udalski2015}
{Udalski}, A., {Szyma{\'n}ski}, M.~K., \& {Szyma{\'n}ski}, G.
  2015{\natexlab{a}}, \actaa, 65, 1

\bibitem[{{Udalski} {et~al.}(2015{\natexlab{b}}){Udalski}, {Yee}, {Gould},
  {Carey}, {Zhu}, {Skowron}, {Koz{\l}owski}, {Poleski}, {Pietrukowicz},
  {Pietrzy{\'n}ski}, {Szyma{\'n}ski}, {Mr{\'o}z}, {Soszy{\'n}ski}, {Ulaczyk},
  {Wyrzykowski}, {Han}, {Calchi Novati}, \& {Pogge}}]{Udalski2015Spitzer}
{Udalski}, A., {Yee}, J.~C., {Gould}, A., {et~al.} 2015{\natexlab{b}}, \apj,
  799, 237

\bibitem[{{Vandorou} {et~al.}(2020{\natexlab{a}}){Vandorou}, {Bennett},
  {Beaulieu}, {Alard}, {Blackman}, {Cole}, {Bhattacharya}, {Bond}, {Koshimoto},
  \& {Marquette}}]{Vandorou2020}
{Vandorou}, A., {Bennett}, D.~P., {Beaulieu}, J.-P., {et~al.}
  2020{\natexlab{a}}, \aj, 160, 121

\bibitem[{{Vandorou} {et~al.}(2020{\natexlab{b}}){Vandorou}, {Bennett},
  {Beaulieu}, {Alard}, {Blackman}, {Cole}, {Bhattacharya}, {Bond}, {Koshimoto},
  \& {Marquette}}]{mb13220b}
{Vandorou}, A., {Bennett}, D.~P., {Beaulieu}, J.-P., {et~al.}
  2020{\natexlab{b}}, \aj, 160, 121

\bibitem[{{Walker}(1995)}]{Walker1995}
{Walker}, M.~A. 1995, \apj, 453, 37

\bibitem[{{Witt} \& {Mao}(1994)}]{WittMao1994}
{Witt}, H.~J. \& {Mao}, S. 1994, \apj, 430, 505

\bibitem[{{Wyrzykowski} \& {Hodgkin}(2012)}]{Wyrzykowski2012}
{Wyrzykowski}, {\L}. \& {Hodgkin}, S. 2012, in IAU Symposium, Vol. 285, New
  Horizons in Time Domain Astronomy, ed. E.~{Griffin}, R.~{Hanisch}, \&
  R.~{Seaman}, 425--428

\bibitem[{{Wyrzykowski} {et~al.}(2016){Wyrzykowski}, {Kostrzewa-Rutkowska},
  {Skowron}, {Rybicki}, {Mr{\'o}z}, {Koz{\l}owski}, {Udalski}, {Szyma{\'n}ski},
  {Pietrzy{\'n}ski}, {Soszy{\'n}ski}, {Ulaczyk}, {Pietrukowicz}, {Poleski},
  {Pawlak}, {I{\l}kiewicz}, \& {Rattenbury}}]{Wyrzykowski2016}
{Wyrzykowski}, {\L}., {Kostrzewa-Rutkowska}, Z., {Skowron}, J., {et~al.} 2016,
  \mnras, 458, 3012

\bibitem[{{Wyrzykowski} \& {Mandel}(2020)}]{WyrzykowskiMandel2020}
{Wyrzykowski}, {\L}. \& {Mandel}, I. 2020, \aap, 636, A20

\bibitem[{{Wyrzykowski} {et~al.}(2020){Wyrzykowski}, {Mr{\'o}z}, {Rybicki},
  {Gromadzki}, {Ko{\l}aczkowski}, {Zieli{\'n}ski}, {Zieli{\'n}ski},
  {Britavskiy}, {Gomboc}, {Sokolovsky}, {Hodgkin}, {Abe}, {Aldi}, {AlMannaei},
  {Altavilla}, {Al Qasim}, {Anupama}, {Awiphan}, {Bachelet},
  {Bak{\i}{\textcommabelow s}}, {Baker}, {Bartlett}, {Bendjoya}, {Benson},
  {Bikmaev}, {Birenbaum}, {Blagorodnova}, {Blanco-Cuaresma}, {Boeva},
  {Bonanos}, {Bozza}, {Bramich}, {Bruni}, {Burenin}, {Burgaz}, {Butterley},
  {Caines}, {Caton}, {Calchi Novati}, {Carrasco}, {Cassan}, {{\v{C}}epas},
  {Cropper}, {Chru{\'s}li{\'n}ska}, {Clementini}, {Clerici}, {Conti}, {Conti},
  {Cross}, {Cusano}, {Damljanovic}, {Dapergolas}, {D'Ago}, {de Bruijne},
  {Dennefeld}, {Dhillon}, {Dominik}, {Dziedzic}, {Erece}, {Eselevich},
  {Esenoglu}, {Eyer}, {Figuera Jaimes}, {Fossey}, {Galeev}, {Grebenev},
  {Gupta}, {Gutaev}, {Hallakoun}, {Hamanowicz}, {Han}, {Handzlik}, {Haislip},
  {Hanlon}, {Hardy}, {Harrison}, {van Heerden}, {Hoette}, {Horne}, {Hudec},
  {Hundertmark}, {Ihanec}, {Irtuganov}, {Itoh}, {Iwanek}, {Jovanovic},
  {Janulis}, {Jel{\'\i}nek}, {Jensen}, {Kaczmarek}, {Katz}, {Khamitov},
  {Kilic}, {Klencki}, {Kolb}, {Kopacki}, {Kouprianov}, {Kruszy{\'n}ska},
  {Kurowski}, {Latev}, {Lee}, {Leonini}, {Leto}, {Lewis}, {Li}, {Liakos},
  {Littlefair}, {Lu}, {Manser}, {Mao}, {Maoz}, {Martin-Carrillo}, {Marais},
  {Maskoli{\={u}}nas}, {Maund}, {Meintjes}, {Melnikov}, {Ment},
  {Miko{\l}ajczyk}, {Morrell}, {Mowlavi}, {Mo{\'z}dzierski}, {Murphy},
  {Nazarov}, {Netzel}, {Nesci}, {Ngeow}, {Norton}, {Ofek},
  {Pak{\v{s}}tien{\.{e}}}, {Palaversa}, {Pandey}, {Paraskeva}, {Pawlak},
  {Penny}, {Penprase}, {Piascik}, {Prieto}, {Qvam}, {Ranc},
  {Rebassa-Mansergas}, {Reichart}, {Reig}, {Rhodes}, {Rivet}, {Rixon},
  {Roberts}, {Rosi}, {Russell}, {Zanmar Sanchez}, {Scarpetta}, {Seabroke},
  {Shappee}, {Schmidt}, {Shvartzvald}, {Sitek}, {Skowron}, {{\'S}niegowska},
  {Snodgrass}, {Soares}, {van Soelen}, {Spetsieri},
  {Stankevi{\v{c}}i{\={u}}t{\.{e}}}, {Steele}, {Street}, {Strobl}, {Strubble},
  {Szegedi}, {Tinjaca Ramirez}, {Tomasella}, {Tsapras}, {Vernet}, {Villanueva},
  {Vince}, {Wambsganss}, {van der Westhuizen}, {Wiersema}, {Wium}, {Wilson},
  {Yoldas}, {Zhuchkov}, {Zhukov}, {Zdanavi{\v{c}}ius}, {Zo{\l}a}, \&
  {Zubareva}}]{WyrzykowskiGaia16aye}
{Wyrzykowski}, {\L}., {Mr{\'o}z}, P., {Rybicki}, K.~A., {et~al.} 2020, \aap,
  633, A98

\bibitem[{{Yee} {et~al.}(2014){Yee}, {Han}, {Gould}, {Skowron}, {Bond},
  {Udalski}, {Hundertmark}, {Monard}, {Porritt}, {Nelson}, {Bozza}, {Albrow},
  {Choi}, {Christie}, {DePoy}, {Gaudi}, {Hwang}, {Jung}, {Lee}, {McCormick},
  {Natusch}, {Ngan}, {Park}, {Pogge}, {Shin}, {Tan}, {{\ensuremath{\mu}}FUN
  Collaboration}, {Abe}, {Bennett}, {Botzler}, {Freeman}, {Fukui}, {Fukunaga},
  {Itow}, {Koshimoto}, {Larsen}, {Ling}, {Masuda}, {Matsubara}, {Muraki},
  {Namba}, {Ohnishi}, {Philpott}, {Rattenbury}, {Saito}, {Sullivan}, {Sumi},
  {Sweatman}, {Suzuki}, {Tristram}, {Tsurumi}, {Wada}, {Yamai}, {Yock},
  {Yonehara}, {MOA Collaboration}, {Szyma{\'n}ski}, {Ulaczyk}, {Koz{\l}owski},
  {Poleski}, {Wyrzykowski}, {Kubiak}, {Pietrukowicz}, {Pietrzy{\'n}ski},
  {Soszy{\'n}ski}, {OGLE Collaboration}, {Bramich}, {Browne}, {Figuera Jaimes},
  {Horne}, {Ipatov}, {Kains}, {Snodgrass}, {Steele}, {Street}, {Tsapras}, \&
  {RoboNet Collaboration}}]{mb13220}
{Yee}, J.~C., {Han}, C., {Gould}, A., {et~al.} 2014, \apj, 790, 14

\bibitem[{{Yoo} {et~al.}(2004){Yoo}, {DePoy}, {Gal-Yam}, {Gaudi}, {Gould},
  {Han}, {Lipkin}, {Maoz}, {Ofek}, {Park}, {Pogge}, {Mu-Fun Collaboration},
  {Udalski}, {Soszy{\'n}ski}, {Wyrzykowski}, {Kubiak}, {Szyma{\'n}ski},
  {Pietrzy{\'n}ski}, {Szewczyk}, {{\.Z}ebru{\'n}}, \& {OGLE
  Collaboration}}]{Yoo2004}
{Yoo}, J., {DePoy}, D.~L., {Gal-Yam}, A., {et~al.} 2004, \apj, 603, 139

\bibitem[{{Zang} {et~al.}(2020){Zang}, {Dong}, {Gould}, {Calchi Novati},
  {Chen}, {Yang}, {Li}, {Mao}, {Alton}, {Brimacombe}, {Carey}, {Christie},
  {Delplancke-Str{\"o}bele}, {Feliz}, {Gaudi}, {Green}, {Hu}, {Jayasinghe},
  {Koff}, {Kurtenkov}, {M{\'e}rand }, {Minev}, {Mutel}, {Natusch}, {Roth},
  {Shvartzvald}, {Sun}, {Vanmunster}, \& {Zhu}}]{Zang2020}
{Zang}, W., {Dong}, S., {Gould}, A., {et~al.} 2020, \apj, 897, 180

\bibitem[{{Zhu} {et~al.}(2017){Zhu}, {Udalski}, {Novati}, {Chung}, {Jung},
  {Ryu}, {Shin}, {Gould}, {Lee}, {Albrow}, {Yee}, {Han}, {Hwang}, {Cha}, {Kim},
  {Kim}, {Kim}, {Kim}, {Lee}, {Park}, {Pogge}, {KMTNet Collaboration},
  {Poleski}, {Mr{\'o}z}, {Pietrukowicz}, {Skowron}, {Szyma{\'n}ski},
  {KozLowski}, {Ulaczyk}, {Pawlak}, {OGLE Collaboration}, {Beichman}, {Bryden},
  {Carey}, {Fausnaugh}, {Gaudi}, {Henderson}, {Shvartzvald}, {Wibking}, \&
  {Spitzer Team}}]{Zhu2017}
{Zhu}, W., {Udalski}, A., {Novati}, S.~C., {et~al.} 2017, \aj, 154, 210

\bibitem[{{Zieli{\'n}ski} {et~al.}(2020){Zieli{\'n}ski}, {Wyrzykowski},
  {Mikolajczyk}, {Rybicki}, \& {Kolaczkowski}}]{Zielinski2020}
{Zieli{\'n}ski}, P., {Wyrzykowski}, L., {Mikolajczyk}, P., {Rybicki}, K., \&
  {Kolaczkowski}, Z. 2020, arXiv e-prints, arXiv:2006.05160

\bibitem[{{Zieli{\'n}ski} {et~al.}(2019){Zieli{\'n}ski}, {Wyrzykowski},
  {Rybicki}, {Ko{\l}aczkowski}, {Bru{\'s}}, \&
  {Miko{\l}ajczyk}}]{Zielinski2019}
{Zieli{\'n}ski}, P., {Wyrzykowski}, {\L}., {Rybicki}, K., {et~al.} 2019,
  Contributions of the Astronomical Observatory Skalnate Pleso, 49, 125

\bibitem[{{Zurlo} {et~al.}(2018){Zurlo}, {Gratton}, {Mesa}, {Desidera}, {Enia},
  {Sahu}, {Almenara}, {Kervella}, {Avenhaus}, {Girard}, {Janson}, {Lagadec},
  {Langlois}, {Milli}, {Perrot}, {Schlieder}, {Thalmann}, {Vigan}, {Giro},
  {Gluck}, {Ramos}, \& {Roux}}]{Zurlo2018}
{Zurlo}, A., {Gratton}, R., {Mesa}, D., {et~al.} 2018, \mnras, 480, 236

\end{thebibliography}

\end{document}